\documentclass[12pt]{article}
\usepackage{geometry} 
\usepackage{braket}
\usepackage{amsmath,amssymb}
\usepackage{graphicx}
\usepackage{epstopdf}
\usepackage[comma,authoryear]{natbib}

\usepackage[hidelinks,breaklinks]{hyperref}
\usepackage[hyphenbreaks]{breakurl}

\DeclareMathOperator{\pr}{Pr}

\begin{document}

\title{Broken arrows: Hardy-Unruh chains \\ and quantum contextuality} 
\author{Michael Janas and Michel Janssen}
\date{\today}

\maketitle

\begin{abstract}
\noindent
\citet{Hardy:1993} and \citet{Unruh:2018} constructed a family of non-maximally entangled states of pairs of particles giving rise to correlations that cannot be accounted for with a local hidden-variable theory. Rather than pointing to violations of some Bell inequality, however, they pointed to clashes with the basic rules of logic. Specifically, they constructed these states and the associated measurement settings in such a way that the outcomes will satisfy a set of two or three conditionals, which we call Hardy-Unruh chains, but not a conditional entailed by this set. Quantum mechanics avoids such broken `if \ldots then \ldots' arrows because it cannot simultaneously assign truth values to all conditionals involved. Measurements to determine the truth value of some preclude measurements to determine the truth value of others. Hardy-Unruh chains thus nicely illustrate quantum contextuality: which variables do and do not get definite values depends on what measurements we decide to perform. We use the framework inspired by \citet{Bub:2016} and \citet{Pitowsky:1989} and developed  in \citet{JanasCuffaroJanssen:2022} to construct and analyze Hardy-Unruh chains in terms of fictitious bananas mimicking the behavior of spin-$\frac12$ particles. 
\end{abstract}

\section{Introduction} \label{Introduction}

The standard way to show that quantum theory allows correlations impossible in classical (more precisely: local hidden-variable) theories is to point to violations of some Bell inequality. A classic example is the violation of the CHSH inequality \citep{ClauserHorneShimonyHolt:1969} by correlations between the outcomes of certain measurements on pairs of photons in a maximally entangled state. An alternative approach is to show that quantum mechanics allows correlations that clash with basic logic. The work by \citet{Hardy:1992,Hardy:1993} and \citet{Unruh:2018} that we will examine in this paper provides intriguing examples of this approach \citep[the most famous example, undoubtedly, is due to][]{GreenbergerHorneZeilinger:1989}. In this approach, at least in principle, one combination of measurement outcomes suffices to rule out a local hidden-variable theory for the relevant quantum correlations whereas in the more familiar approach we need to consider the statistics of many outcomes.    

\citet{Hardy:1993} constructed a family of non-maximally entangled two-particle states and concomitant measurement settings such that the measurement outcomes satisfy two conditionals, but not a third, which would seem to be a direct consequence of the first two. Schematically, 
\begin{equation}
A \rightarrow C, \quad B \rightarrow D, \quad (A \, \& \, \, B) \not\to (C \, \& \, \, D).
\label{Hardy chain 12} 
\end{equation}
This is what is known as Hardy's paradox.

Inspired by Hardy, \citet{Unruh:2018} constructed a family of states and settings such that the outcomes satisfy three conditionals, but not a fourth, which would seem to follow directly from the first three on the basis of the transitivity of the `if \ldots\ then' relation. Schematically,
\begin{equation}
A \rightarrow B, \quad B \rightarrow C, \quad C \rightarrow D, \quad A \not\to D.
\label{Unruh chain 12} 
\end{equation}

Such broken `if \ldots\ then \ldots' arrows are allowed in quantum mechanics for the same reason that violations of Bell inequalities are. Local hidden-variable theories simultaneously assign truth values to propositions $A$, $B$, $C$ and $D$ above. Quantum mechanics does not. To assign truth values to all four propositions, one would simultaneously have to measure observables represented by non-commuting operators. These Hardy-Unruh chains of conditionals---as we will call the sets of conditionals in Eqs.\ (\ref{Hardy chain 12}) and \ (\ref{Unruh chain 12})---thus illustrate quantum contextuality: which observables do and do not get definite values depends on what measurements we decide to perform.

In this paper, we use the framework inspired by  \citet{Bub:2016} and \citet{Pitowsky:1989} and developed in \citet{JanasCuffaroJanssen:2022} to construct and analyze these Hardy-Unruh chains. In Section \ref{Preliminaries}, we review the elements we need from our book. In Sections \ref{Hardy states} and \ref{Hardy-Unruh states}, we construct the states and measurement settings giving rise to the broken arrows in Eqs.\ (\ref{Hardy chain 12}) and (\ref{Unruh chain 12}). In Section \ref{Geometrical representation of the correlations found with Hardy-Unruh states}, we examine the relation between these broken arrows and violations of the relevant Bell inequality, which, as we will see, is a special case of the CHSH inequality. On the basis of this analysis, we conclude, in Section \ref{Conclusion}, that broken arrows and violations of Bell inequalities are just slightly different but ultimately equivalent ways of bringing out quantum contextuality.

\section{Preliminaries} \label{Preliminaries} 

In \emph{Understanding Quantum Raffles} \citep{JanasCuffaroJanssen:2022}, inspired by \emph{Bananaworld} \citep{Bub:2016}, we used the imagery of peeling and tasting fictitious bananas mimicking the measurement of spin components of (half-)integer spin particles. We modified Bub's banana-peeling scheme to tighten the analogy between our bananas and particles with spin. In this paper, as in most of our book, we focus on bananas mimicking the behavior of spin-$\frac12$ particles.\footnote{The figures in this section are all based on figures in \citet{JanasCuffaroJanssen:2022}. 
As in the book (see p.\ xvii), the two of us focus on pedagogy (Janssen) and polytopes (Janas) and leave the philosophy to Cuffaro (see his contribution to this issue). Fittingly, the title of our paper comes from a Neil Young song on Buffalo Springfield's sophomore album. A song with the same title appears on the first solo album of another Canadian artist, Robbie Robertson (1943--2023).}  

Imagine picking a pair of such bananas, connected at the stem, from a particular species of banana tree, breaking them apart and giving one to Alice and one to Bob. Alice and Bob then choose a peeling direction, i.e., a direction in which they are required to hold their banana while peeling it. When done peeling, they take a bite to determine whether their banana tastes yummy or nasty, these being the only two possible tastes these bananas can acquire upon being tasted. Readers put off by our \emph{Bananaworld} conceit can replace (i) bananas by spin-$\frac12$ particles; (ii) species of banana trees by states in which we prepare pairs of such particles (though we will also talk about pairs of bananas in particular quantum states); (iii) peeling directions (or peelings for short) by orientations of Du Bois (or Stern-Gerlach) magnets; (iv) the actual peeling by sending particles through a Du Bois magnet; (v) tasting by having a particle hit a screen behind the magnet with a photo-emulsion; and (vi) yummy and nasty by spin up and spin down. 

\begin{figure}[h]
 \centering
   \includegraphics[width=2.1in]{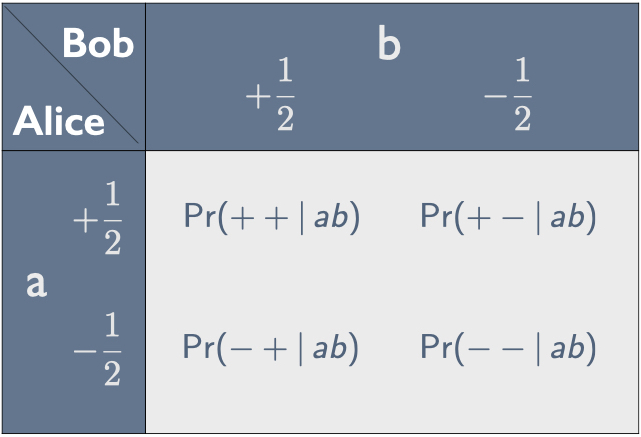} 
   \caption{\small{Correlation array for Alice peeling $a$ and Bob peeling $b$.}} 
   \label{Correlation-array-cell-prob}
\end{figure}

Suppose Alice peels $a$ and Bob peels $b$. The correlations between the tastes they find, which persist no matter how far apart they are, can be represented by a \emph{correlation array} (see Fig.\ \ref{Correlation-array-cell-prob}). In analogy with the values $+\frac12\hbar$ and $-\frac12\hbar$ for spin up and spin down (where $\hbar$ is Planck's constant divided by $2\pi$), we assign the numerical values $+\frac12$ and $-\frac12$ to the tastes yummy and nasty in some appropriate units. Unless we need these values to calculate expectation values, we will simply use $+$ for yummy and $-$ for nasty. The four entries in the correlation array give the probabilities 
of the four possible outcomes for this combination of peelings.

\begin{figure}[h]
 \centering
   \includegraphics[width=2.1in]{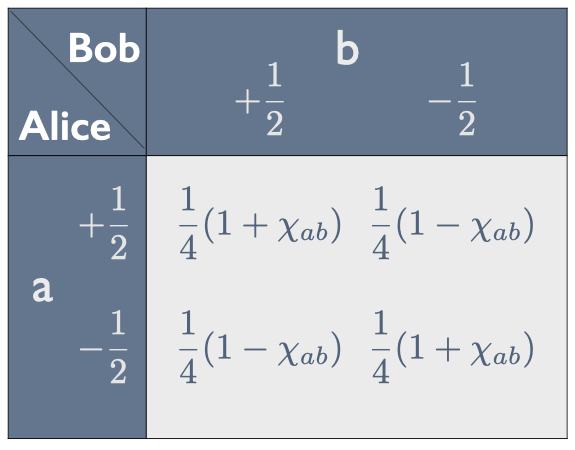} 
   \caption{\small{Parametrization of correlation array in Fig.\ \ref{Correlation-array-cell-prob} given property 1.}} 
   \label{Correlation-array-cell-chi}
\end{figure}

For now, we restrict our attention to species of banana trees (but this does \emph{not} include the species giving rise to Hardy-Unruh chains) on which bananas grow in pairs such that the correlations between their tastes have two special properties:
\begin{enumerate}
\vspace{-.2cm}
\item No matter what peelings Alice and Bob use, the probability of them finding yummy or nasty is always $\frac12$. 
\vspace{-.2cm}
\item If Alice and Bob use the same peeling, they always find opposite tastes.
\vspace{-.2cm}
\end{enumerate}
Property 1 means that the entries in both rows and both columns of the correlation array in Fig.\ \ref{Correlation-array-cell-prob} add up to $\frac12$. In that case, as shown in Fig.\ \ref{Correlation-array-cell-chi}, the correlation array can be fully  characterized by the parameter $-1 \le \chi_{ab} \le +1$, with $\chi_{ab} = -1$ if the peelings $a$ and $b$ are the same (property 2). 

\begin{figure}[h]
 \centering
   \includegraphics[width=3.5in]{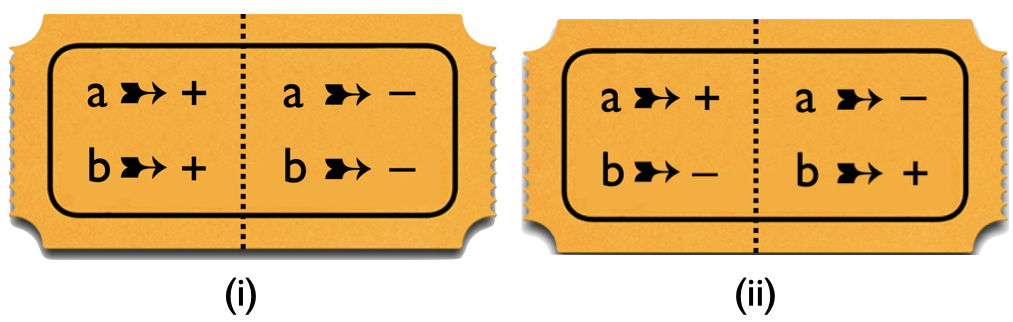} 
   \caption{\small{Raffle tickets.}}
   \label{raffle-tickets-ab}
\end{figure}

We can simulate these correlations for any value of $\chi_{ab}$ with the kind of raffle introduced in \citet[sec.\ 2.5]{JanasCuffaroJanssen:2022} as a model for local hidden-variable theories. In this case, the raffle consists of a basket with a mix of the two types of tickets shown in Fig.\ \ref{raffle-tickets-ab}, with the tastes of both bananas for both peelings printed on them. We draw tickets from this basket, tear them in half along the perforation indicated by the dashed line, and randomly give one half to Alice and one half to Bob. That the values for $a$ and $b$ on the two sides of the ticket are opposite takes care of property 2. That we randomly decide which half goes to Alice and which half to Bob takes care of property 1.

A raffle that exclusively has tickets of type (i) will give a perfect anti-correlation between Alice's result for $a$ and Bob's result for $b$. In that case, the entries on the diagonal in Figs.\ \ref{Correlation-array-cell-prob}--\ref{Correlation-array-cell-chi} are $0$ while the off-diagonal ones are $\frac12$. So for tickets of type (i), $\chi_{ab} = -1$. A raffle that exclusively has tickets of type (ii) will give a perfect correlation. In that case, the off-diagonal entries in Figs.\ \ref{Correlation-array-cell-prob}--\ref{Correlation-array-cell-chi} are $0$ and those on the diagonal are $\frac12$. So for tickets of type (i), $\chi_{ab} = 1$. To simulate the correlation in Fig.\ \ref{Correlation-array-cell-chi} for arbitrary values of $\chi_{ab}$ we need a raffle with $\frac12 (1 - \chi_{ab}) \times 100 \%$ tickets of type (i) and $\frac12 (1 + \chi_{ab}) \times 100 \%$ tickets of type (ii).

It turns out that, for all values between $-1$ and $+1$, $\chi_{ab}$ is the (\emph{Pearson}) \emph{correlation coefficient} of the variables $A_a$ and $B_b$, the taste Alice finds when peeling $a$ and the taste Bob finds when peeling $b$.\footnote{In \citet[pp.\ 72--73]{JanasCuffaroJanssen:2022}, $\chi_{ab}$ is defined as an \emph{anti}-correlation coefficient. The extra minus sign affects Table \ref{values of chi}, Fig.\ \ref{elliptope} and Eqs.\ (\ref{chi matrix})--(\ref{elliptope inequality}) below.\label{extra minus sign}} The correlation coefficient of two stochastic variables $X$ and $Y$ is defined as the covariance, $\mathrm{Cov}(XY) \equiv \langle (X - \langle X \rangle) (Y - \langle Y \rangle)\rangle$, divided by the standard deviations, $\sigma_X$ and $\sigma_Y$, the square roots of the variances, $\langle (X - \langle X \rangle)^2 \rangle$ and $\langle (Y - \langle Y \rangle)^2 \rangle$.  What simplifies matters in the case of the variables $A_a$ and $B_b$ is that they are \emph{balanced}, i.e., their two possible values are each other's opposite and these two values are equiprobable \citep[p.\ 68]{JanasCuffaroJanssen:2022}. This means that their expectation values, $\langle A_a \rangle$ and $\langle B_b \rangle$, vanish and that the correlation coefficient is given by
 \begin{equation}
\rho_{A_a B_b} = \frac{\langle A_a B_b\rangle}{\sqrt{\langle A_a^2 \rangle} \sqrt{\langle B_b^2 \rangle}}.
\label{def correlation coefficient}
\end{equation}
Inspection of the correlation arrays in Figs.\ \ref{Correlation-array-cell-prob}--\ref{Correlation-array-cell-chi} tells us that
\begin{eqnarray}
\langle A_a B_b\rangle &\!\!\!=\!\!\!& {\textstyle \frac14} \Big( \pr(+\!+\!|ab) + \pr(-\!-\!|ab) \Big) \, - \,  {\textstyle \frac14} \Big(\pr(+\!-\!|ab) + \pr(-\!+\!|ab) \Big) \nonumber \\[.2cm]
 &\!\!\!=\!\!\!&  {\textstyle \frac14} \cdot {\textstyle \frac12} \big(1+ \chi_{ab}\big) \, - \,  {\textstyle \frac14} \cdot {\textstyle \frac12} \big(1 - \chi_{ab}\big) = {\textstyle \frac14} \, \chi_{ab};
 \label{covariance formula}
\end{eqnarray}
that
\begin{equation}
\langle A_a^2 \rangle = {\textstyle \frac14} \Big( \pr(+\!+\!|ab) + \pr(+\!-\!|ab)\Big)  +  {\textstyle \frac14}  \Big( \pr(-\!+\!|ab) +  \pr(-\!-\!|ab) \Big) =  {\textstyle \frac14};
\label{variance}
\end{equation}
and that, similarly, $\langle B_b^2 \rangle = \frac14$. Substituting these results into Eq.\ (\ref{def correlation coefficient}), we see that the correlation coefficient is indeed equal to the parameter characterizing the correlation in Fig.\ \ref{Correlation-array-cell-chi}:
\begin{equation}
\rho_{A_a B_b} = \frac{\frac14 \, \chi_{ab}}{\frac12 \cdot \frac12} = \chi_{ab}.
\label{rho = chi}
\end{equation}

As noted above, unless $\chi_{ab} = \pm 1$, we need a mix of tickets to simulate the correlation array in Fig.\ \ref{Correlation-array-cell-chi} with one of our raffles. With our quantum bananas we can produce this correlation array for arbitrary values $-1 < \chi_{ab} < 1$ with pairs of bananas in the familiar fully entangled singlet state, \emph{but with different combinations of peeling directions}. Using the bases $\{ |\pm \rangle_a \}$ and $\{ |\pm \rangle_b \}$ of eigenvectors of the operators representing the observables `taste when peeled in the $a$-direction' and `taste when peeled in the $b$-direction' for the one-banana Hilbert space to construct bases for the two-banana Hilbert space, we can write the singlet state as:
\begin{equation}
\ket{\psi_{\mathrm{singlet}}} = \frac{1}{\sqrt{2}} \Big( \ket{+-}_{aa} - \ket{-+}_{aa} \Big)
= \frac{1}{\sqrt{2}} \Big( \ket{+-}_{bb} - \ket{-+}_{bb} \Big),
\label{singlet state}
\end{equation}
where $\ket{+-}_{aa}$ etc.\ is shorthand for the tensor product $|+\rangle_a \otimes |-\rangle_a$ etc.

\begin{figure}[h]
 \centering
   \includegraphics[width=2.5in]{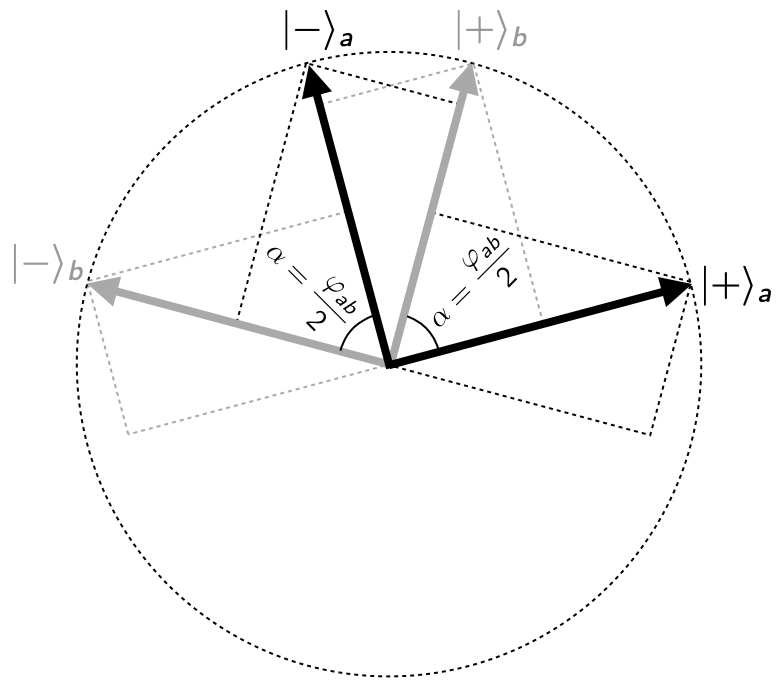} 
   \caption{\small{Eigenvectors for `taste when peeled in the $a$-direction' and `taste when peeled in the $b$-direction' in the one-banana Hilbert space, where $\varphi_{ab} =2\alpha$ is the angle between the peeling directions $a$ and $b$.}} 
   \label{Hardy-Unruh-bananas}
\end{figure}

The relation between the $a$-basis and the $b$-basis is illustrated in Fig.\ \ref{Hardy-Unruh-bananas}. The angle $\alpha$ between these pairs of eigenvectors is equal to half the angle $\varphi_{ab}$ between the peeling directions $a$ and $b$. The transformation from the $b$-basis to the $a$-basis is given by:
\begin{equation}
\begin{array}{c}
 |+ \rangle_a = \cos{\alpha} \, |+ \rangle_b - \sin{\alpha} \, |- \rangle_b,  \\[.4cm]
 |- \rangle_a = \sin{\alpha} \, |+ \rangle_b + \cos{\alpha} \, |- \rangle_b \,;
\end{array}
\label{a in terms of b}
\end{equation}
its inverse by:
\begin{equation}
\begin{array}{c}
 |+ \rangle_b = \cos{\alpha} \, |+ \rangle_a + \sin{\alpha} \, |- \rangle_a,  \\[.4cm]
|- \rangle_b = - \sin{\alpha} \, |+ \rangle_a + \cos{\alpha} \, |- \rangle_a.
\end{array}
\label{b in terms of a}
\end{equation}

\begin{figure}[h]
\centering
    \includegraphics[width=2.5in]{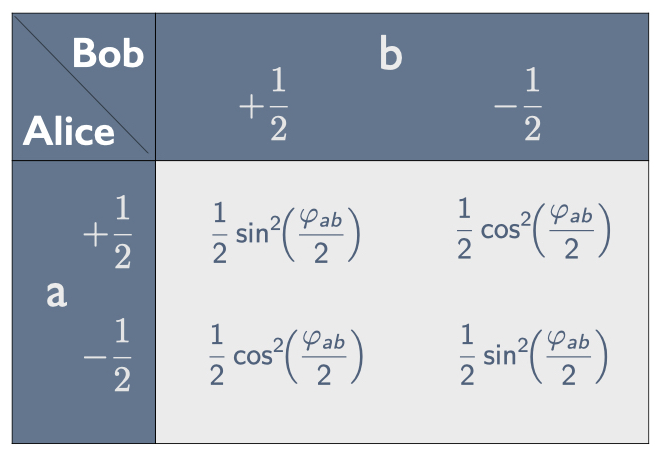}
    \vspace{-.2cm}
 \caption{\small{Correlation array for taste-and-peel experiment with bananas in the singlet state.}}
 \label{CA-singlet-cell}
\end{figure}

To find the probabilities of the various combinations of outcomes when Alice peels $a$ and Bob peels $b$, we use these transformation equations to write the singlet state in the $ab$-basis:
\begin{equation}
\ket{\psi_{\mathrm{singlet}}} = \frac{1}{\sqrt{2}} \Big( \sin{\alpha}  \ket{++}_{ab} 
\, + \, \cos{\alpha}  \ket{+-}_{ab}
\, - \, \cos{\alpha}  \ket{-+}_{ab} 
\, + \, \sin{\alpha}  \ket{--}_{ab} 
 \Big).
 \label{singlet state ab basis}
\end{equation}
The Born rule tells us that the probabilities of finding the various combinations of tastes when Alice peels $a$ and Bob peels $b$ are given by the squares of the coefficients of the corresponding terms of the singlet state in the $ab$-basis. Recalling that $\alpha = \varphi_{ab}/2$, we thus arrive at the correlation array in Fig.\ \ref{CA-singlet-cell}.

Using this correlation array to calculate the correlation coefficient (see \ Eq.\ (\ref{def correlation coefficient})), we find:
\begin{equation}
\rho_{A_a B_b} = \frac{\frac14 \cdot \sin^2{\! \displaystyle{\Big( \frac{\varphi_{ab}}{2}\Big)}} - \frac14 \cdot \cos^2{\! \displaystyle{\Big( \frac{\varphi_{ab}}{2}\Big)}}}{\frac12 \cdot \frac12} = -\cos{\varphi_{ab}}.
\label{chi = - cos}
\end{equation}
We saw earlier (see Eq.\ (\ref{rho = chi})) that $\rho_{A_a B_b}$ is equal to the parameter $\chi_{ab}$ characterizing the correlation array in Fig.\ \ref{Correlation-array-cell-chi}. With the appropriate choice of peeling directions we can thus obtain this correlation array for any value $-1 \le \chi_{ab} \le 1$ with the appropriate measurements on the same quantum state, whereas we needed a mix of tickets to obtain this correlation array with one of our raffles.

In \citet{JanasCuffaroJanssen:2022}, we used the tools introduced above to analyze the correlations found in an experimental setup due to \citet{Mermin:1981} in which Alice and Bob peel and taste bananas in the singlet state choosing between three different peeling directions, $a$, $b$ and $c$. The correlations between the tastes found by Alice and Bob in this Mermin setup can be represented by a 3$\times$3 correlation array with cells of the form shown in Fig.\ \ref{Correlation-array-cell-chi} with $\chi_{ab} = - \cos{\varphi_{ab}}$ etc.\ (see Fig.\ \ref{CA-singlet-cell} and Eq.\ (\ref{chi = - cos})). 

Because of the symmetry of the singlet state, the cells of the correlation array on one side of the diagonal ($ab$, $ac$ and $bc$) are the same as those on the other side ($ba$, $ca$ and $cb$). In the cells on the diagonal we have a perfect anti-correlation (if $a =b$, $\varphi_{ab} = 0$ and  $\chi_{ab} = -1$). A correlation array for this Mermin setup can thus be characterized by the correlation coefficients for half of its off-diagonal cells, $\chi_{ab}$, $\chi_{ac}$ and $\chi_{bc}$, all three taking on values between $-1$ and $+1$. 

Inspired by \citet{Pitowsky:1989}, we used these coefficients as coordinates of a point in a 2$\times$2$\times$2 cube, the \emph{non-signaling polytope} ($\mathcal{P}$) for the Mermin setup (see Fig.\ \ref{elliptope}). The part of $\mathcal{P}$ allowed by quantum mechanics is called the \emph{quantum convex set} ($\mathcal{Q}$); the part allowed by local hidden-variable theories the \emph{local polytope} ($\mathcal{L}$).\footnote{See \citet{Gohetal:2018} and \citet{Leetal:2023} for interesting recent work in this tradition, to which Tsirelson made important early contributions, as illustrated, for instance, by Fig.\ 2 in \citet[p.\ 3]{Tsirelson:1993}. Five years earlier, at the beginning of a section entitled ``Representations of extremal correlations,'' he already noted: ``As one can easily see, the set Cor$(m,n)$ of all quantum realized $m \times n$ correlation matrices \ldots\ is a closed, bounded, centrally symmetric, convex body in the space of $m \times n$ matrices'' \citep[p.\ 562]{Tsirelson:1987}.} 

\begin{figure}[h]
\centering
    \includegraphics[width=5.5in]{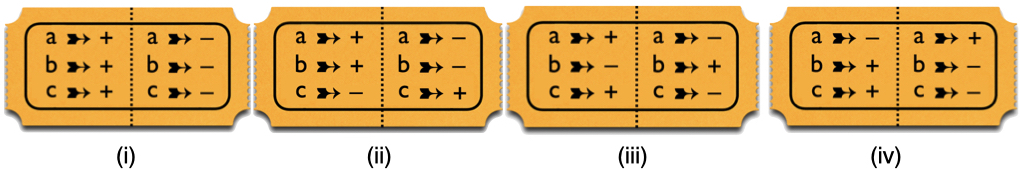}
    \vspace{-.2cm}
 \caption{\small{Tickets for a raffle meant to simulate the correlation array for the Mermin setup.}}
 \label{raffle-tickets-3set2out-i-thru-iv-row} 
\end{figure}  

We derive the inequalities defining $\mathcal{L}$ and $\mathcal{Q}$ in this case. As our model for a local hidden-variable theory, we use a raffle with a mix of the four types of tickets shown in Fig.\ \ref{raffle-tickets-3set2out-i-thru-iv-row}. The values of the correlation coefficients for raffles with only one type of ticket can be read directly off that ticket. For example, if the values for $a$ and $b$ on opposite sides of the ticket are the same, $\chi_{ab} = 1$; if they are opposite, $\chi_{ab} = -1$. Table \ref{values of chi} collects the values of $\chi_{ab}$, $\chi_{ac}$ and $\chi_{bc}$ for ticket types (i)--(iv).

\begin{table}[h]
\centering
\begin{tabular}{|c||c|c|c|}
\hline
ticket & \quad $\chi_{ab}$ \quad & \quad $\chi_{ac}$ \quad & \quad $\chi_{bc}$ \quad \\[.1cm] 
\hline
 (i) & $-1$ & $-1$ & $-1$ \\[.2cm]
 (ii) & $-1$ & $+1$ & $+1$ \\[.2cm]
 (iii) & $+1$ & $-1$ & $+1$ \\[.2cm]
(iv) & $+1$ & $+1$ & $-1$ \\
 \hline
\end{tabular} 
\caption{Values of the anti-correlation coefficients for raffles with just one of the four types of tickets shown in Fig.\ \ref{raffle-tickets-3set2out-i-thru-iv-row}.}
\label{values of chi}
\end{table} 

The correlations produced by raffles with just one of these four ticket-types are represented by the vertices that are labeled (i) through (iv) in the non-signaling cube in Fig.\ \ref{elliptope}. The \emph{local polytope} ($\mathcal{L}$) for the Mermin setup is the tetrahedron formed by these four vertices. 

\begin{figure}[h]
\centering
    \includegraphics[width=3.5in]{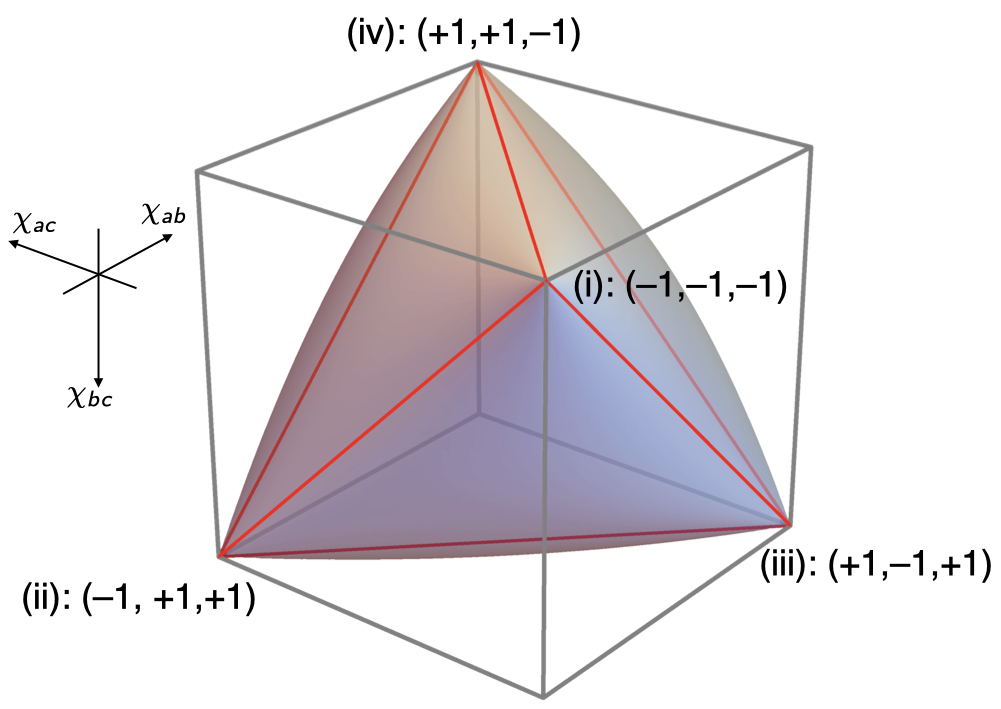}
 \caption{\small{The non-signaling polytope ($\mathcal{P}$), the quantum convex set ($\mathcal{Q}$) and the local polytope ($\mathcal{L}$) for the Mermin setup.}}
 \label{elliptope}  
\end{figure}

The Bell inequality for the Mermin setup corresponds to one of the four facets of the tetrahedron, the one with the vertices (ii), (iii) and (iv). The pair of inequalities associated with this facet, which can be read off Table \ref{values of chi}, is:
\begin{equation}
-3 \le \chi_{ab} + \chi_{ac} + \chi_{bc} \le 1.
\label{Mermin inequality}
\end{equation}
This is the direct analogue of the CHSH inequality, the Bell inequality for a setup involving four rather than three different peelings, with Alice peeling $a$ or $b$ and Bob peeling $a^\prime$ or $b^\prime$ \citep[cf.\ Eq.\ (\ref{CHSH ineqs}) below and][Ch.\ 5]{JanasCuffaroJanssen:2022}. To fully characterize the local polytope for the Mermin setup, we need three more pairs of inequalities like the ones in  Eq.\ (\ref{Mermin inequality}), corresponding to the other three facets of the tetrahedron in Fig.\ \ref{elliptope}.

To find the \emph{quantum convex set} ($\mathcal{Q}$) for the Mermin setup, we consider the 3$\times$3 matrix formed by the correlation coefficients characterizing the nine cells of its correlation array. Using that $\chi_{ab} = -\cos{\varphi_{ab}} = - \vec{e}_a \cdot \vec{e}_b$ etc.\ (where $\vec{e}_a$ and $\vec{e}_b$ are unit vectors in the peeling directions $a$ and $b$), we can write this \emph{correlation matrix} as:   
\begin{equation}
\chi
\equiv  
\begin{pmatrix}
\; -1 \; & \; \chi_{ab} \; & \;  \chi_{ac} \; \\[.2cm]
\; \chi_{ba} & \; -1 \; & \;  \chi_{bc} \; \\[.2cm]
 \; \chi_{ca} \; & \; \chi_{cb} \; & \;  -1 
\end{pmatrix}
=
- \begin{pmatrix}
\; \vec{e}_a \cdot \vec{e}_a \; & \; \vec{e}_a \cdot \vec{e}_b \; & \;  \vec{e}_a \cdot \vec{e}_c \; \\[.2cm]
\; \vec{e}_b \cdot \vec{e}_a & \; \vec{e}_b \cdot \vec{e}_b \; & \;  \vec{e}_b \cdot \vec{e}_c \; \\[.2cm]
 \; \vec{e}_c \cdot \vec{e}_a \; & \; \vec{e}_c \cdot \vec{e}_b \; & \;  \vec{e}_c \cdot \vec{e}_c 
\end{pmatrix}.
\label{chi matrix}
\end{equation}
This is (minus) a \emph{Gram matrix}, which has the property that its determinant cannot be negative: $- \det{\chi} \ge 0$. This gives us the constraint we are looking for:
\begin{equation}
1 - \chi_{ab}^2 - \chi_{ac}^2 - \chi_{bc}^2 - 2 \, \chi_{ab} \, \chi_{ac} \, \chi_{bc} \ge 0.
\label{elliptope inequality}
\end{equation}
This non-linear inequality defines the elliptope representing the quantum convex set ($\mathcal{Q}$) for the Mermin setup in Fig.\ \ref{elliptope}.\footnote{ Taking a slice of this figure by setting one of the $\chi$'s to zero, we obtain the Vitruvian-man-like cartoon in \citet[p.\ 107, Fig.\ 5.2]{Bub:2016} for $\mathcal{P}$, $\mathcal{Q}$ and $\mathcal{L}$ in an arbitrary setup.}  

We now have all the ingredients we need from \citet{JanasCuffaroJanssen:2022} to analyze the correlations found with Hardy and Hardy-Unruh states.

\section{Hardy states} \label{Hardy states}

 \citet{Hardy:1993} cooked up a family of two-particle states, each member with its own combination of measurements to be performed on it, to illustrate the apparent breakdown of basic logic in quantum mechanics (see Eq.\ (\ref{Hardy chain 12})). We construct the states for a branch of this family in \emph{Bananaworld}, in which Alice and Bob both use the same pair of peelings $a$ and $b$. As we will see when we turn to the intimately related Hardy-Unruh family of states, other members of the Hardy family involve Alice and Bob using different pairs of peelings, which we will label $(a,b)$ and $(a',b')$, respectively. 

\subsection{Hardy chain of conditionals} \label{Hardy chain of conditionals}

Hardy states have four special properties that translate into corresponding properties of the correlations between the tastes found by Alice, peelings $a$ or $b$, and Bob, peeling $a'$ or $b'$, which can but do not have to be the same as $a$ and $b$.

\begin{enumerate}
\item  
\vspace{-.2cm}
There is no $\ket{+-}$ component in the $ba'$-basis. So if Alice peels $b$ and finds +, then Bob will also find + when he peels $a'$. Schematically: $A_{b_+} \rightarrow B_{a^\prime_+}$.\footnote{Of course, this property of the state also implies $B_{a^\prime_-} \rightarrow A_{b_-}$, but this conditional is not part of the Hardy chain.}
\item  
\vspace{-.2cm}
There is no $\ket{-+}$ component in the $ab'$-basis. So if Bob peels $b'$ and finds +, then Alice will also find + when he peels $a$. Schematically: $B_{b^\prime_+} \rightarrow A_{a_+}$.
\item 
\vspace{-.2cm}
There is no $\ket{++}$ component in the $aa'$-basis. So if Alice peels $a$ and Bob peels $a'$, they cannot both find +. Schematically: we cannot have $A_{a_+} \& \, B_{a^\prime_+}$.
\item 
\vspace{-.2cm}
There is both a $\ket{++}$ and a $\ket{+-}$ component in the $bb'$-basis. So if Alice peels $b$ and Bob peels $b'$, it is possible for both of them to find +. Schematically: we can have $A_{b_+} \& \, B_{b^\prime_+}$.
\end{enumerate}
\begin{figure}[h]
\vspace{-.2cm}
 \centering
   \includegraphics[width=2.5in]{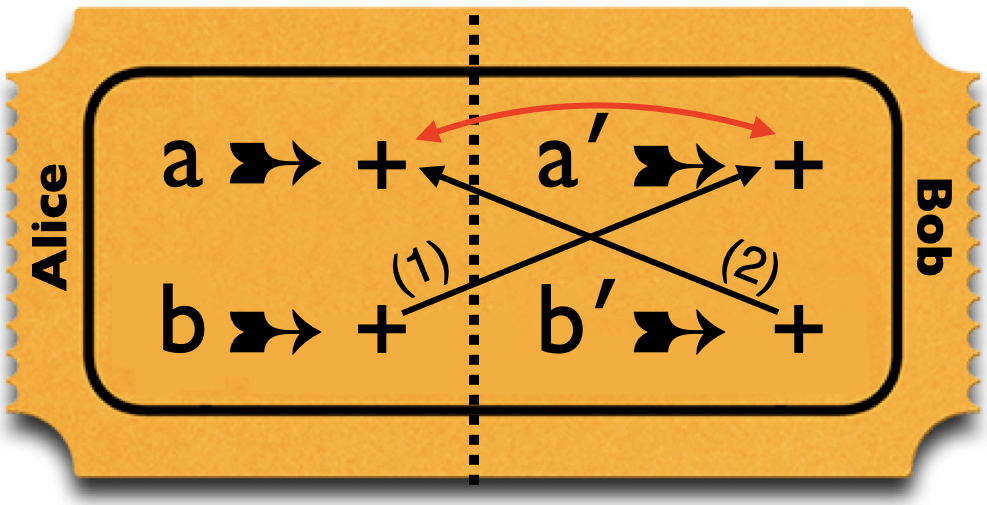} 
   \caption{\small{Conflicting demands on the design of a ticket for a raffle simulating the correlations found in measurements on Hardy states.}}
   \label{H-state-raffle-ticket-design}
\end{figure}

These four properties place contradictory demands on the design of tickets for a raffle simulating these correlations. This is illustrated in Fig.\ \ref{H-state-raffle-ticket-design}. Since Alice and Bob use different pairs of peelings, the left side of the ticket always goes to Alice and the right side to Bob.
Because of property 4, our raffle must contain \emph{some} tickets with $+$ for both $b$ and $b'$. Because of properties 1 and 2, such tickets must also have $+$ for both $a$ and $a'$. However, because of property 3, our raffle is not allowed to contain \emph{any} such tickets! 

Following \citet[p.\ 1666]{Hardy:1993}, we can bring out the problem in a slightly different way  \citep[see also][p.\ 34]{KwiatHardy:2000}. The conditionals $A_{b_+} \rightarrow B_{a^\prime_+}$ (property 1) and $B_{b^\prime_+} \rightarrow A_{a_+}$ (property 2) entail the composite conditional
\begin{equation}
\big( A_{b_+} \, \mathrm{and} \, B_{b^\prime_+} \big) \rightarrow \big( A_{a_+} \, \mathrm{and} \, B_{a^\prime_+} \big).
\label{combined conditional}
\end{equation}
But this conditional is false: it is possible for the antecedent to be true (property 4) and the consequent to be false (property 3). Quantum mechanics avoids the broken arrow in Eq.\ (\ref{combined conditional}) by not allowing truth values to be assigned simultaneously to antecedent and consequent. The same pair of bananas cannot be peeled and tasted twice: Alice cannot peel hers both $a$ and $b$, Bob cannot peel his both $a'$ and $b'$.

\subsection{Constructing Hardy states} \label{Constructing Hardy states}

We construct a branch of the family of Hardy states in \emph{Bananaworld} with $a = a'$ and $b = b'$. Members of this branch can be labeled by the angle $\alpha$, which is half the angle $\varphi_{ab}$ between the peeling directions $a$ and $b$. The angle $\alpha$ thus runs from $0$ to $\pi/2$. We start with property 3: the state has no $\ket{++}$ component in the $aa$-basis:
\begin{equation}
\ket{\psi_H(\alpha)} = N(\alpha) \Big( - \sin{\alpha} \ket{+-}_{aa} + \cos{\alpha}  \ket{--}_{aa} - \sin{\alpha} \ket{-+}_{aa} \Big),
\label{H state aa basis}
\end{equation}
where the factor
\begin{equation}
N(\alpha) = \frac{1}{\sqrt{1 + \sin^2{\!\alpha}}}
\label{H state normalization}
\end{equation}
normalizes the state. The coefficients of the three components in the $aa$-basis were chosen with malice aforethought. Given our choice of peeling $b$ to go with peeling $a$, these coefficients ensure that $\ket{\psi_{H}(\alpha)}$ has properties 1 and 2. Combining the first and the second term on the right-hand side of Eq.\ (\ref{H state aa basis}) and using Eq.\ (\ref{b in terms of a}), the transformation from the $a$- to the $b$-basis, we can write $\ket{\psi_{H}(\alpha)}$ as
\begin{eqnarray}
\ket{\psi_{H}(\alpha)} &\!\!\! = \!\!\! & N(\alpha) \Big( \Big\{ -  \sin{\alpha} \, |+\rangle_a + \cos{\alpha} \, |-\rangle_a \Big\} \otimes |-\rangle_a - \sin{\alpha} \ket{-+}_{aa} \Big)
\nonumber\\[.2cm]
&\!\!\! = \!\!\! & N(\alpha) \Big(  \ket{--}_{ba}  -  \sin{\alpha} \ket{-+}_{aa}  \Big),
\label{H state ba basis partial}
\end{eqnarray}
which shows that $\ket{\psi_{H}(\alpha)}$ has no $\ket{+-}$ component in the $ba$-basis (property 1). Combining the second and the third term on the right-hand side of Eq.\ (\ref{H state aa basis}), we can also write $\ket{\psi_{H}(\alpha)}$ as
\begin{eqnarray}
\ket{\psi_{H}(\alpha)} &\!\!\! = \!\!\! & N(\alpha) \Big( - \sin{\alpha} \ket{+-}_{aa} + |-\rangle_a \otimes \Big\{\cos{\alpha} \, |-\rangle_a  -  \sin{\alpha} \, |+\rangle_a \Big\} \Big)
\nonumber\\[.2cm]
&\!\!\! = \!\!\! & N(\alpha) \Big(  - \sin{\alpha} \ket{+-}_{aa} + \ket{--}_{ab} \Big),
\label{H state ab basis partial}
\end{eqnarray}
which shows that $\ket{\psi_{H}(\alpha)}$ has no $\ket{-+}$ component in the $ab$-basis (property 2).

Finally, starting from Eq.\ (\ref{H state ba basis partial}) (but we could also have started from Eq.\ (\ref{H state ab basis partial})) and using Eq.\ (\ref{a in terms of b}), the transformation from the $b$- to the $a$-basis, we can write $\ket{\psi_{H}(\alpha)}$ in the $bb$-basis:
\begin{eqnarray}
\ket{\psi_{H}(\alpha)} & \!\!\! = \!\!\! & N(\alpha) \Big( \, |-\rangle_b \otimes \Big\{\! \sin{\alpha} \,  |+ \rangle_b   +  \cos{\alpha} \, |- \rangle_b \Big\}  
\nonumber \\[.1cm]
&& \hspace{1cm} -  \sin{\alpha} \, \Big\{\! \sin{\alpha} \,  |+ \rangle_b   +  \cos{\alpha} \, |- \rangle_b \Big\} \otimes \Big\{\! \cos{\alpha} \,  |+ \rangle_b   -  \sin{\alpha} \, |- \rangle_b \Big\} \Big) \nonumber \\[.25cm]
 & \!\!\! = \!\!\! & N(\alpha) \Big(\! - \sin^2{\!\alpha} \cos{\alpha} \ket{++}_{bb}  \, + \, \sin^3{\!\alpha} \,  \ket{+-}_{bb}   \nonumber \\[.1cm]
 && \hspace{1.5cm} + \, \sin^3{\!\alpha} \ket{-+}_{bb}  \, + \, \cos{\alpha} \, (1 + \sin^2{\alpha})  \ket{--}_{bb} \Big).
\label{H state bb basis}
\end{eqnarray} 
This shows that $\ket{\psi_{H}(\alpha)}$ has both a $\ket{++}$ and a $\ket{+-}$ component  in the $bb$-basis (property 4). 

To construct a correlation array for the results of Alice and Bob peeling pairs of bananas in the Hardy state, we need $\ket{\psi_{H}(\alpha)}$ in the $aa$-, $ab$-, $ba$- and $bb$-basis. Eqs.\ (\ref{H state aa basis}) and Eqs.\ (\ref{H state bb basis}) give the state in the $aa$- and $bb$-basis, respectively. Starting from Eqs.\ (\ref{H state ba basis partial}) and (\ref{H state ab basis partial}), we find the state in the $ba$- and $ab$-basis, respectively:
\begin{eqnarray}
\ket{\psi_{H}(\alpha)} & \!\!\! = \!\!\! & N(\alpha) \Big(  \ket{--}_{ba}  -  \sin{\alpha} \, \Big\{ \sin{\alpha} \ket{+}_b + \cos{\alpha} \ket{-}_b \Big\} \otimes \ket{+}_a \Big) 
 \nonumber \\[.2cm]
 && N(\alpha) \Big( \ket{--}_{ba} \, - \, \sin^2{\!\alpha} \ket{++}_{ba} \, - \, \sin{\alpha} \cos{\alpha} \ket{-+}_{ba} \Big)
 \label{H state ba basis} \\[.4cm]
\ket{\psi_{H}(\alpha)} & \!\!\! = \!\!\! & N(\alpha) \Big(  - \sin{\alpha} \ket{+}_a \otimes \Big\{ \sin{\alpha} \ket{+}_b + \cos{\alpha} \ket{-}_b \Big\} + \ket{--}_{ab} \Big) 
\nonumber \\[.2cm]
 && N(\alpha) \Big( - \sin^2{\!\alpha} \ket{++}_{ab} \, - \, \sin{\alpha} \cos{\alpha} \ket{+-}_{ab} \, + \, \ket{--}_{ab} \Big)
 \label{H state ab basis} 
\end{eqnarray}

\begin{figure}[h]
\centering
    \includegraphics[width=3.8in]{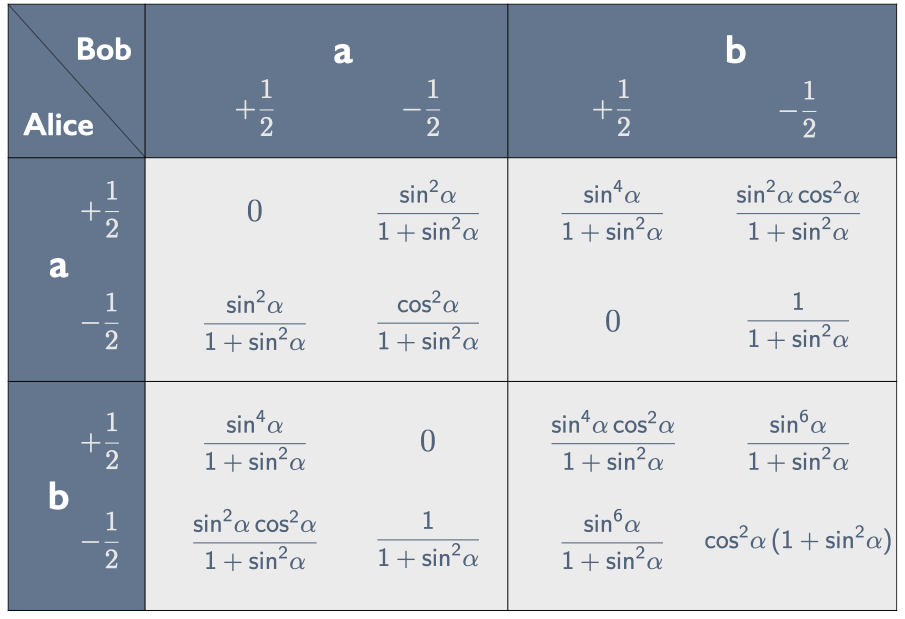}
 \caption{\small{Correlation array for the tastes $\pm\frac12$ of pairs of bananas in the Hardy state in Eqs.\ (\ref{H state aa basis})--(\ref{H state ab basis}), both peeled $a$ or $b$ by Alice and Bob.}}
 \label{H-correlation-array}  
\end{figure}

Using the Born rule, we can read off all probabilities entering into the correlation array in Fig.\ \ref{H-correlation-array} from  Eqs.\ (\ref{H state aa basis}) and (\ref{H state bb basis})--(\ref{H state ab basis}). One readily checks that (i) in each of the four cells the four entries sum to 1 and (ii) in each row and column the sum of the first two entries is equal to the sum of the last two (though verifying this for the last row involves some tedious algebra). Property (ii) guarantees that the correlation is non-signaling: the marginal probabilities $\mathrm{Pr}(A_{a_\pm})$ and $\mathrm{Pr}(A_{b_\pm})$ for Alice do not depend on the peeling chosen by Bob and vice versa \citep[cf.][pp.\ 25--26]{JanasCuffaroJanssen:2022}.\footnote{This is true for any compound system with a Hilbert space that is the tensor product of the Hilbert spaces of its components. We sketch a simple proof of this property for the situation at hand, which can readily be adapted to the general case. The probability that Alice finds $+$ when she peels $a$ and Bob peels $b$ can be written as
$$
\mathrm{Pr}\Big( A_+ \big| A_a, B_b, \hat{\rho} \Big) = \mathrm{Pr}\Big( A_+ \& \, B_+ \big| A_a, B_b, \hat{\rho} \Big) + \mathrm{Pr}\Big( A_+ \& \, B_- \big| A_a, B_b, \hat{\rho} \Big),
$$ 
where $\hat{\rho}$ is the density operator characterizing the quantum statistical ensemble under consideration. We are considering the uniform ensemble of pairs of quantum bananas in the state $\ket{\psi_{HU}(\alpha)}$, so $\hat{\rho}$ is simply equal to $\ket{\psi_{HU}(\alpha)} \bra{\psi_{HU}(\alpha)}$, the projection operator onto that state, but the same argument works for any ensemble of pairs of quantum bananas in any state. Using the trace formula for the Born rule for both terms, we can write the right-hand side as
$$
\mathrm{Tr}\Big( \hat{\rho}  \, \Big( \hat{P}_{1_{a_+}} \!\!\otimes \hat{P}_{2_{b_+}} \Big) \Big)
+  \mathrm{Tr}\Big( \hat{\rho}  \, \Big( \hat{P}_{1_{a_+}} \!\!\otimes \hat{P}_{2_{b_-}} \Big) \Big).
$$
where $\hat{P}_{1_{a_\pm}}$ and $\hat{P}_{2_{b_\pm}}$ are the projection operators in the Hilbert spaces of the bananas of Alice (subscript 1) and Bob (subscript 2) onto the eigenvectors in the $a$- and $b$-bases. Using basic properties of the trace operation and the tensor product, we can rewrite this expression as
$$
\mathrm{Tr}\Big( \hat{\rho} \, \Big( \hat{P}_{1_{a_+}} \!\!\otimes \big[ \hat{P}_{2_{b_+}} + \hat{P}_{2_{b_-}} \big] \Big) \Big).
$$
Using the completeness of the $b$-basis to set $\hat{P}_{2_{b_+}} + \hat{P}_{2_{b_-}} = \hat{1}_2$, we arrive at 
$$
\mathrm{Pr}\Big( A_+ \big| A_a, B_b, \hat{\rho} \Big)  = \mathrm{Tr}\Big( \hat{\rho} \, \Big( \hat{P}_{1_{a_+}} \!\!\otimes \hat{1}_2 \Big) \Big).
$$
Had Bob chosen $a$, we would have arrived at the exact same result, using the completeness of the $a$-basis. This shows that this marginal probability is indeed independent of the peeling Bob chooses.\label{tensor product gives non-signaling}}
 
We can read the properties of Hardy states listed in Section \ref{Hardy chain of conditionals} (with $a'=a$ and $b'=b$) directly off the correlation array in Fig.\ \ref{H-correlation-array}. That $\pr(+\!-\!|ba)=0$ gives the conditional $A_{b_+} \rightarrow  B_{a_+}$ (property 1). That $\pr(-\!+\!|ab)=0$ likewise  gives the conditional $B_{b_+} \rightarrow  A_{a_+}$ (property 2). The $aa$ and $bb$ cells show that the composite conditional is false: $(A_{b_+} \& \, B_{b_+}) \not\to (B_{a_+} \& \, A_{a_+})$. That $\pr(+\!+\!|bb) \neq 0$ means that we can have $A_{b_+} \& \, B_{b_+}$ (property 4); that $\pr(+\!+\!|aa) = 0$ means we cannot have $B_{a_+} \& \, A_{a_+}$ (property 3).\footnote{The correlation array in Fig.\ \ref{H-correlation-array}---like the one in Fig.\ \ref{HU-correlation-array} below---belongs to ``Class 3a: Three blocks [cells in our terminology] having one zero each'' in the helpful classification of $2\!\times\!2$ correlation arrays by \citet{Chenet al:2023}.}    

We cannot simulate this correlation array with one of our raffles because a raffle that gives the $0$'s in the $aa$, $ab$ and $ba$ cells must also give a $0$ for the $++$ entry in the $bb$ cell (cf.\ the ticket in Fig.\ \ref{H-state-raffle-ticket-design}). Finding one instance (or a few to allow for experimental error) of Alice and Bob both finding $+$ when peeling $b$ would thus rule out a local hidden-variable theory capable of producing this correlation array. 

\subsection{Hardy states between maximally entangled and product states} \label{Hardy states between maximally entangled and product states}

The Hardy states $\ket{\psi_{H}(\alpha)}$ in Eqs.\ (\ref{H state aa basis}) and (\ref{H state bb basis})--(\ref{H state ab basis}) result in the broken arrow in Eq.\ (\ref{combined conditional}) unless $\alpha = 0$ or $\alpha = \pi/2$. What happens in those two cases?

For $\alpha = 0$, $N(\alpha) =1$ and Eqs.\ (\ref{H state aa basis}) and (\ref{H state ba basis partial})--(\ref{H state bb basis}) reduce to
\begin{equation}
|\psi_{H}(0) \rangle = \ket{--}
\label{H state alpha 0}
\end{equation}
in all four bases ($aa$, $ab$, $ba$ and $bb$). The state thus becomes a product state (a property independent of the basis we choose) and the correlations it gives rise to can easily be simulated with one of our raffles. 

For $\alpha = \pi/2$, $N = 1/\sqrt{2}$ and Eqs.\ (\ref{H state aa basis}) and (\ref{H state bb basis}) reduce to
\begin{equation}
\ket{\psi_{H}(\textstyle{\frac\pi2}} = - \frac{1}{\sqrt{2}} \, \Big( \ket{+-}_{aa} + \ket{-+}_{aa} \Big)
= \frac{1}{\sqrt{2}} \, \Big(  \ket{+-}_{bb} + \ket{-+}_{bb} \Big).
\label{Hardy pi/2 max entangled}
\end{equation}
This is a maximally entangled state (cf.\ the singlet state in Eq.\ (\ref{singlet state})). That the expansion in the $aa$-basis differs by a minus sign from the expansion in the $bb$ basis reflects that, if $\alpha = \pi/2$, $|+\rangle_b = |-\rangle_a$ and $|-\rangle_b = -|+\rangle_a$. It may sound paradoxical  that we can simulate the correlations generated by this maximally entangled state whereas for the non-maximally entangled Hardy states we cannot. Remember, however, that for $\alpha = \varphi_{ab}/2 = \pi/2$, the peeling directions $a$ and $b$ are exactly opposite. This case (like the case when the peeling directions are the same: $\alpha =0$) can easily be simulated with one of our raffles. 

\citet{KwiatHardy:2000} consider the special case that $\cos{\alpha} = \sqrt{2/5}$ and $\sin{\alpha} = \sqrt{3/5}$, which means that $\alpha \approx 51^{\mathrm o}$. In that case (see Eq.\ (\ref{H state normalization})),
\begin{equation}
N(\alpha) \, \sin{\alpha} = \frac{\sin{\alpha}}{\sqrt{1 + \sin^2{\!\alpha}}} = \sqrt{\frac38}, \quad N(\alpha) \, \cos{\alpha} = \frac{\cos{\alpha}}{\sqrt{1 + \sin^2{\!\alpha}}} = \frac12,
\end{equation}
and Eq.\ (\ref{H state aa basis}) becomes:\footnote{Kwiat and Hardy present their example is terms of quantum cakes rather than quantum bananas. Our conditions 1--4 are their conditions 2, 2$^\prime$, 3 and 1, respectively \citep[p.\ 34]{KwiatHardy:2000}. Instead of `taste when peeled $a$' and `taste when peeled $b$' (with values + for yummy and $-$ for nasty), they introduce the variables `taste' (with values $G$ and $B$ for `good' and `bad') and `rising of batter' (with values for $R$ and $N$ for `risen' and `not risen'). The corresponding orthonormal bases, $\{ |G\rangle, |B\rangle \}$ and $\{ |R\rangle, |N\rangle \}$, are related via (cf.\ Eq.\ (\ref{a in terms of b}) for $\alpha \approx 51^{\mathrm o}$):
$$
 |G \rangle = {\textstyle \sqrt{\frac25}} \, |R  \rangle - {\textstyle \sqrt{\frac35}} \, | N \rangle, \quad |B \rangle = {\textstyle \sqrt{\frac35} } \, |R \rangle + {\textstyle \sqrt{\frac25}} \, |N \rangle.
$$
Using the $\{ |G\rangle, |B\rangle \}$ basis for the Hilbert space of both quantum cakes, they write the state in Eq.\ (\ref{KH state in aa basis}) as (with $L$ and $R$ for `left' or `Lucien' and `right' or `Ricardo' instead of our 1 or 2):
$$
|\psi\rangle = {\textstyle \frac12} \, |B_L \rangle |B_R\rangle - {\textstyle \sqrt{\frac38}} \, \Big[ |B_L \rangle |G_R\rangle  + |G_L \rangle |B_R\rangle \Big]
$$
\citep[see][p.\ 35, Appendix]{KwiatHardy:2000}.\label{bananas and cakes}}
\begin{equation}
{\textstyle |\psi_{H}(\alpha \approx 51^{\mathrm o}) \rangle =  - \sqrt{\frac38} \ket{+-}_{aa}  + \frac12 \, \ket{--}_{aa} - \sqrt{\frac38} \, \ket{-+}_{aa} }.
\label{KH state in aa basis}
\end{equation}
The Born rule tells us that the probability of Alice and Bob both finding $+$ when both are peeling $b$ is equal to the square of the coefficient of $\ket{++}$ of $ |\psi_{H}(\alpha) \rangle$  in the $bb$-basis. For $\alpha \approx 51^{\mathrm o}$, this coefficient is
\begin{equation}
- \frac{\sin^2{\!\alpha} \cos{\alpha}}{\sqrt{1 + \sin^2{\!\alpha}}} = - \frac{\frac35 \cdot \sqrt{\frac25}}{\sqrt{\frac85}} = -\frac{3}{5\cdot2}. 
\label{maximum violation}
\end{equation}
Hence $\pr(+\!+\!|bb) = 0.09$ \citep[cf.][p.\ 34]{KwiatHardy:2000}. As \citet[p.\ 885]{Mermin:1994} notes, this is ``only a shade [$\approx 0.0002$] less than the maximum possible'' for the square of the expression on the left-hand side of Eq.\ (\ref{maximum violation}). Mermin gives this maximum as $(2/(1 + \sqrt{5}))^5$ (ibid., p.\ 884); \citet[p.\ 1667]{Hardy:1993} as $\frac12(5 \sqrt{5} -11)$.

This is as far as we will take our analysis of the Hardy family of states. In the next two sections, we will scrutinize the intimately related Hardy-Unruh family more closely, especially the dependence of the correlations generated by a branch of this family on the angle $\alpha$ parametrizing this branch. 

\section{Hardy-Unruh states} \label{Hardy-Unruh states} 

Inspired by Hardy, \citet{Unruh:2018} cooked up a family of states providing an even more striking example than \citet{Hardy:1993} and \citet{KwiatHardy:2000} of the apparent breakdown of basic logic in quantum mechanics  (see Eq.\ (\ref{Unruh chain 12})). Since the Unruh family will turn out to be the same as the Hardy family, we will call these states Hardy-Unruh rather than Unruh states. Our discussion in this section mirrors but will be more general than our discussion in Section \ref{Hardy states}. 

\subsection{Hardy-Unruh chain of conditionals} \label{Hardy-Unruh chain of conditionals}

Hardy-Unruh states have four special properties that translate into corresponding properties of the correlations between the tastes found by Alice, peeling $a$ or $b$, and Bob, peeling $a'$ or $b'$, which can but do not have to be the same as $a$ and $b$:
\begin{enumerate}
\item 
There is no $\ket{+-}$ component in the $ab'$-basis. So if Alice peels $a$ and finds +, Bob will also find + when he peels $b'$.  Schematically: $A_{a_+} \rightarrow B_{b'_+}$.
\item 
\vspace{-.2cm}
There is no $\ket{-+}$ component in the $bb'$-basis. So if Bob peels $b'$ and finds +, Alice will also find + when she peels $b$. Schematically: $B_{b'_+} \rightarrow A_{b_+}$.
\item 
\vspace{-.2cm}
There is  no $\ket{+-}$ component in the $ba'$-basis. So if Alice peels $b$ and finds +, Bob will also find + when he peels $a'$.  Schematically: $A_{b_+} \rightarrow B_{a'_+}$.
\item 
\vspace{-.2cm}
There is both a $\ket{++}$ and a $\ket{+-}$ component in the $aa'$-basis. So if Alice peels $a$ and finds +, Bob might find + or $-$ when he peels $a'$. Schematically: it's possible to have $A_{a_+} \, \& \, \,B_{a'_-}$.
\vspace{-.2cm}
\end{enumerate}

\begin{figure}[h]
 \centering
   \includegraphics[width=2.5in]{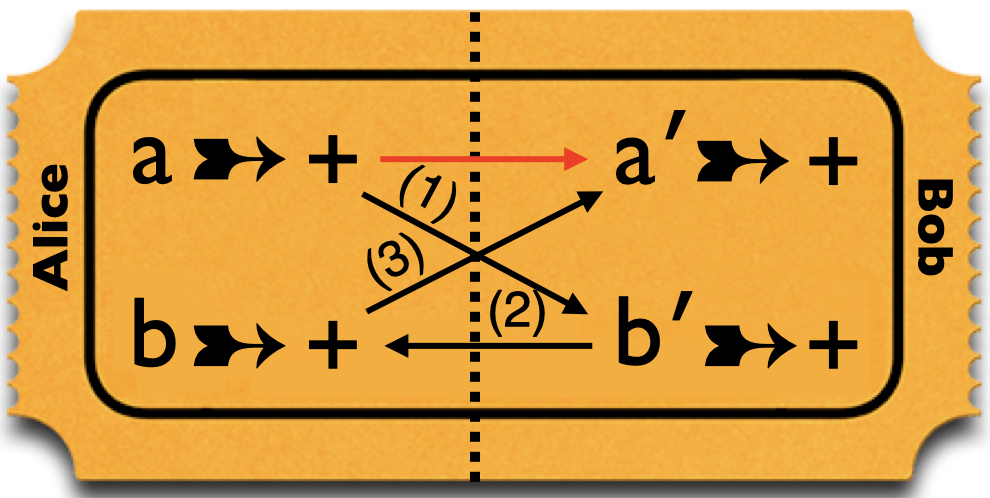} 
   \caption{\small{Conflicting demands on the design of a ticket for a raffle simulating the correlations found in measurements on Hardy-Unruh states.}}
   \label{HU-state-raffle-ticket-design}
\end{figure}

These four properties place contradictory demands on the design of tickets for a raffle simulating these correlations. This is illustrated in Fig.\ \ref{HU-state-raffle-ticket-design}. Because of the conditionals in 1--3, a ticket with $+$ for $a$, must have $+$ for all four entries. However, property 4 requires our raffle to contain at least \emph{some} tickets with three $+$'s (for $a$, $b'$ and $b$) and one $-$ (for $a'$). 

As Fig.\ \ref{HU-state-raffle-ticket-design} illustrates, the conditionals expressing properties 1--3 can be combined into the chain of conditionals
\begin{equation}
A_{a_+} \rightarrow B_{b'_+} \rightarrow A_{b_+} \rightarrow B_{a'_+}.
\label{chain of conditionals}
\end{equation}
Yet $A_{a_+} \not\to B_{a'_+}$: it is possible for the antecedent of this conditional to be true and the consequent to be false (property 4).  As with the broken arrow in the Hardy case (cf.\ Eq.\ (\ref{combined conditional})), quantum mechanics avoids the problem by not allowing truth values to be assigned  simultaneously to $A_{a_+}$ and $A_{b_+}$ or to $B_{a^\prime_+}$ and $B_{b^\prime_+}$. The same banana cannot be peeled and tasted twice.

\subsection{Constructing Hardy-Unruh states} \label{Constructing Hardy-Unruh states}

Our construction of the family of Hardy-Unruh states follows the same pattern as our construction of a branch of the family of Hardy states in Eqs.\ (\ref{H state aa basis})--(\ref{H state bb basis}). We start by making sure that the state has property 2, i.e., that it has no $\ket{-+}$ component in the $bb'$-basis:
\begin{equation}
\ket{\psi_{HU}(u,v,w)} = N \Big( u \ket{++}_{bb'} - v \ket{+-}_{bb'} - w \ket{--}_{bb'} \Big),
\label{HU state bb'}
\end{equation}
where $u$, $v$ and $w$ are arbitrary complex numbers and the normalization factor is given by:
\begin{equation}
N \equiv \frac{1}{\sqrt{|u|^2 + |v|^2 + |w|^2}}.
\label{Nuvw}
\end{equation}
This shows how generic these Hardy-Unruh states are. We can construct them by starting from a state orthogonal to any two-particle state that can be written in the form $\ket{-+}$ in an orthonormal basis $\{ \ket{\pm}_b \otimes \ket{\pm}_{b'} \}$ of eigenvectors for some pair of peelings $b$ and $b'$ for Alice and Bob. This is as true for Hardy states as for Hardy-Unruh states.
 
As we did in Eqs.\ (\ref{H state ba basis partial}) and (\ref{H state ab basis partial}), we can group the three terms on the right-hand side of Eq.\ (\ref{HU state bb'}) in two different ways:
\begin{eqnarray}
\ket{\psi_{HU}(u,v,w)} & \!\!\! = \!\!\! & N \Big( \ket{+}_b \otimes \Big\{ u \ket{+}_{b'} - v \ket{-}_{b'} \Big\} - w \ket{--}_{bb'} \Big)
\label{HU state ba' basis partial prep} \\[.2cm]
& \!\!\! = \!\!\! & N \Big( u \ket{++}_{bb'} - \Big\{ v \ket{+}_b + w \ket{-}_b \Big\} \otimes \ket{-}_{b'}  \Big)
\label{HU state ab' basis partial prep}
\end{eqnarray}
Now choose peeling $a$ to go with peeling $b$ such that the corresponding eigenvectors are:
\begin{equation}
\begin{array}{ccc}
\ket{-}_a & \!\!\! = \!\!\! & \displaystyle{\frac{1}{\sqrt{|v|^2 + |w|^2}}} \, \Big( v \ket{+}_b + w \ket{-}_b \Big) \\[.6cm]
\ket{+}_a & \!\!\! = \!\!\! &  \displaystyle{\frac{1}{\sqrt{|v|^2 + |w|^2}}} \, \Big( \overline{w} \ket{+}_b - \overline{v} \ket{-}_b \Big)
\end{array}
\label{a in terms of b vw}
\end{equation}
(where bars denote complex conjugates); and choose peeling $a'$ to go with peeling $b'$ such that the corresponding eigenvectors are:
\begin{equation}
\begin{array}{ccc}
\ket{+}_{a'} & \!\!\! = \!\!\! & \displaystyle{\frac{1}{\sqrt{|u|^2 + |v|^2}}} \, \Big( u \ket{+}_{b'} - v \ket{-}_{b'} \Big) \\[.6cm]
\ket{-}_{a'} & \!\!\! = \!\!\! &  \displaystyle{\frac{1}{\sqrt{|u|^2 + |v|^2}}} \, \Big( \overline{v} \ket{+}_{b'} + \overline{u} \ket{-}_{b'} \Big).
\end{array}
\label{a' in terms of b' uv}
\end{equation}
If $\{ \ket{\pm}_b \}$ and $\{ \ket{\pm}_{b'} \}$ are orthonormal bases, then $\{ \ket{\pm}_a \}$ and $\{ \ket{\pm}_{a'} \}$ are too.

Using Eqs.\ (\ref{a in terms of b vw})--(\ref{a' in terms of b' uv}), we can write Eqs.\ (\ref{HU state ba' basis partial prep})--(\ref{HU state ab' basis partial prep}) as
\begin{eqnarray}
\ket{\psi_{HU}(u,v,w)} & \!\!\! = \!\!\! & N \Big( \sqrt{|u|^2 + |v|^2} \, \ket{++}_{ba'} \, - \, w \ket{--}_{bb'} \Big),
\label{HU state ba' basis partial} \\[.2cm]
& \!\!\! = \!\!\! & N \Big( u \ket{++}_{bb'} \, - \, \sqrt{|v|^2 + |w|^2} \, \ket{--}_{ab'}  \Big).
\label{HU state ab' basis partial}
\end{eqnarray}
Eqs.\ (\ref{HU state ba' basis partial})--(\ref{HU state ab' basis partial}) show that $\ket{\psi_{HU}(u,v,w)}$ has no $\ket{+-}$ component in either the $ba'$- or the $ab'$-basis (properties 1 and 3). Finally, using Eqs.\ (\ref{a in terms of b vw})--(\ref{a' in terms of b' uv}) to write $\ket{\psi_{HU}(u,v,w)}$ in the $aa'$-basis, one can verify that $\ket{\psi_{HU}(u,v,w)}$ has both a $\ket{++}$  and a $\ket{+-}$ component in the $aa'$-basis (property 4).

We will only verify this last property for the branch of the family we will focus on in the rest of this paper. The chain in Eq.\ (\ref{chain of conditionals}) already leads to a broken arrow if Alice and Bob use the same peelings $a$ and $b$. In that case, $v=\overline{v}$ and $u=\overline{w}$ in Eq.\ (\ref{a in terms of b vw})--(\ref{a' in terms of b' uv}). We take $u$ and $w$ to be real as well and set:
\begin{equation}
u = w = \cos{\alpha}, \quad v = \sin{\alpha},
\label{uvw cos sin}
\end{equation}
where as before, $0 < \alpha < \pi/2$  is half the angle $\varphi_{ab}$ between the peeling directions $a$ and $b$. With this choice for $(u,v,w)$, Eq.\ (\ref{HU state bb'}) becomes:
\begin{equation}
\ket{\psi_{HU}(\alpha)} = N(\alpha) \Big( \cos{\alpha} \ket{++}_{bb}  -   \sin{\alpha} \,  \ket{+-}_{bb}  -  \cos{\alpha} \, \ket{--}_{bb} \Big),
\label{HU state bb}
\end{equation}
with the normalization factor (cf.\ Eq.\ (\ref{Nuvw}))
 \begin{equation}
N(\alpha) = \frac{1}{\sqrt{1 + \cos^2{\!\alpha}}}.
\end{equation}
Note the similarity to the Hardy state  in Eq.\ (\ref{H state aa basis}). Like $\ket{\psi_{HU}(\alpha)}$, $\ket{\psi_H(\alpha)}$ corresponds to a more general state, $\ket{\psi_{H}(u,v,w)}$, of the same form as $\ket{\psi_{HU}(u,v,w)}$ in Eq.\ (\ref{HU state bb'}).

With the values for $(u,v,w)$ in Eq.\ (\ref{uvw cos sin}), Eqs.\ (\ref{a in terms of b vw})--(\ref{a' in terms of b' uv}) both reduce to Eq.\ (\ref{a in terms of b}) for the transformation from the $b$- to the $a$-basis. Using the inverse transformation, Eq.\ (\ref{b in terms of a}), and substituting the values of $u$, $v$ and $w$ in Eq.\ (\ref{uvw cos sin}) into Eq.\ (\ref{HU state ba' basis partial}), we find $\ket{\psi_{HU}(\alpha)}$ in the $ba$-basis:
\begin{eqnarray}
|\psi_{HU}(\alpha) \rangle & \!\!\! = \!\!\! & N(\alpha) \Big( \ket{++}_{ba} -  \cos{\alpha} \ket{-}_b  \otimes  \Big(\! - \sin{\alpha} \ket{+}_a +  \cos{\alpha} \ket{-}_a \Big) \Big) 
 \nonumber \\[.2cm]
 & \!\!\! = \!\!\! & N(\alpha) \Big( \ket{++}_{ba}  +
\cos{\alpha} \sin{\alpha} \ket{-+}_{ba}   - 
\cos^2{\!\alpha} \ket{--}_{ba} \Big). 
\label{HU state ba} 
\end{eqnarray} 
Eq.\ (\ref{HU state ab' basis partial}) similarly allows us to find $\ket{\psi_{HU}(\alpha)}$ in the $ab$-basis:
\begin{eqnarray}
|\psi_{HU}(\alpha) \rangle & \!\!\! = \!\!\! & N(\alpha) \Big(  \cos{\alpha}  \Big( \! \cos{\alpha} \, |+ \rangle_a   +  \sin{\alpha} \, |- \rangle_a \Big) \otimes |+\rangle_b  -  \ket{--}_{ab} \Big) 
\nonumber  \\[.2cm]
 & \!\!\! = \!\!\! & N(\alpha) \Big( \cos^2{\!\alpha} \ket{++}_{ab}  +  \cos{\alpha} \sin{\alpha} \ket{+-}_{ab} - \ket{--}_{ab} \Big).
\label{HU state ab}  
\end{eqnarray} 
 Finally, starting from Eq.\ (\ref{HU state ab' basis partial})---but we could have started from Eq.\ (\ref{HU state ba' basis partial}) instead---we find $\ket{\psi_{HU}(\alpha)}$ in the $aa$-basis:
\begin{eqnarray}
\ket{\psi_{HU}(\alpha)} & \!\!\! = \!\!\! & N(\alpha) \Big( \cos{\alpha} \ket{++}_{bb} - \ket{--}_{ab} \Big)  \nonumber \\[.15cm]
 & \!\!\! = \!\!\! & N(\alpha) \Big( \cos{\alpha} \, \big\{ \cos{\alpha}  \ket{+}_a + \sin{\alpha} \ket{-}_a \big\} \otimes \big\{ \cos{\alpha}  \ket{+}_a + \sin{\alpha} \ket{-}_a \big\} \nonumber \\
 & & \hspace{5.7cm} - \; \ket{-}_a \otimes \big\{ \cos{\alpha} \ket{-}_a - \sin{\alpha} \ket{+}_a \big\} \Big)  \nonumber \\[.15cm]
& \!\!\! = \!\!\! & N(\alpha) \Big( \cos^3{\!\alpha}  \ket{++}_{aa} + 
 \cos^2{\!\alpha} \sin{\alpha}  \ket{+-}_{aa} \nonumber \\
  && \hspace{2cm} + \, \sin{\alpha} (1+\cos^2{\!\alpha})  \ket{-+}_{aa}  -  \cos^3{\!\alpha}  \ket{--}_{aa}  \Big).
\label{HU state aa}
\end{eqnarray} 
This confirms that $\ket{\psi_{HU}(\alpha)}$ has both a $\ket{++}$ and a $\ket{+-}$ component in the $aa$-basis (property 4). 

\begin{figure}[h]
\centering
    \includegraphics[width=3.8in]{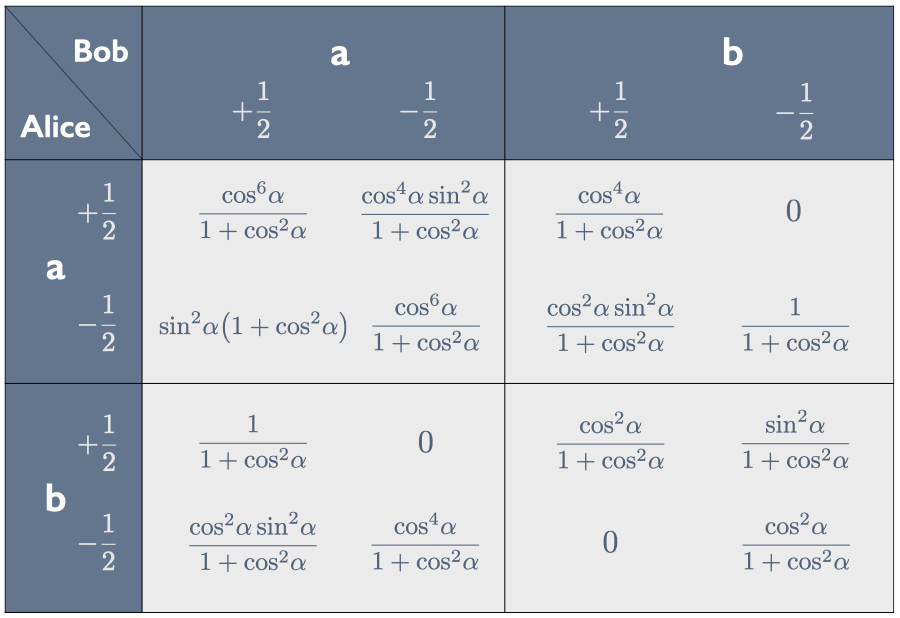}
 \caption{\small{Correlation array for the tastes $\pm\frac12$ of pairs of bananas in the Hardy-Unruh state in Eqs.\ (\ref{HU state bb})--(\ref{HU state aa}), both peeled $a$ or $b$ by Alice and Bob.}}
 \label{HU-correlation-array}  
\end{figure}

We can use Eqs.\ (\ref{HU state bb})--(\ref{HU state aa}) for $\ket{\psi_{HU}(\alpha)}$ in the $aa$-, $ab$-, $ba$- and $bb$-bases to construct the correlation array in Fig.\ \ref{HU-correlation-array}. As with the correlation array for the Hardy state $\ket{\psi_{H}(\alpha)}$ in Fig.\ \ref{H-correlation-array}, we can read properties 1--4 listed in Section \ref{Hardy-Unruh chain of conditionals} (with $a'=a$ and $b'=b$) of the Hardy-Unruh state $\ket{\psi_{HU}(\alpha)}$  directly off the correlation array in Fig.\ \ref{HU-correlation-array}. That $\pr(+\!-\!|ab) = \pr(-\!+\!|bb) = \pr(+\!-|ba) = 0$ translates into the conditionals $A_{a_+} \rightarrow B_{b_+}$, $B_{b_+} \rightarrow A_{b_+}$  and $A_{b_+} \rightarrow B_{a_+}$ (properties 1--3). That $\pr(+\!-|aa) \neq 0$ results in the broken arrow: $A_{a_+} \not\to B_{a_+}$. 

Relabeling peelings and tastes for Alice and Bob and replacing $\alpha$ by $\frac\pi2 - \alpha$, we can turn the correlation array in Fig.\ \ref{HU-correlation-array}  for the Hardy-Unruh state $\ket{\psi_{HU}(\alpha)}$ into the correlation array in Fig.\ \ref{H-correlation-array} for the Hardy state $\ket{\psi_{H}(\frac\pi2 - \alpha)}$. Specifically, we need to make four changes in these correlation arrays to turn one into the other:
\begin{itemize}
\item Switch $\sin{\alpha}$ and $\cos{\alpha}$.
\vspace{-.2cm}
\item Change $(a_\pm,b_\pm)$ to $(b_\pm, a_\mp)$ for Alice.
\vspace{-.2cm}
\item Change $(a_\pm,b_\pm)$ to $(b_\mp, a_\pm)$ for Bob.
\vspace{-.2cm}
\item Switch rows and columns to get back to the standard format with labels in the order $(a_+, a_-, b_+, b_-)$ for both Alice and Bob.
\end{itemize}
As we have seen, the correlation arrays in Figs.\ \ref{H-correlation-array} and \ref{HU-correlation-array} capture the defining properties of Hardy and Hardy-Unruh states listed in Sections \ref{Hardy chain of conditionals} and \ref{Hardy-Unruh chain of conditionals}, respectively. That one can be obtained from the other through the simple expedient of relabeling rows and columns and switching sines and cosines shows that these states are all members of one and the same family.

We cannot simulate the correlation array in Fig.\ \ref{HU-correlation-array} with one of our raffles because a raffle that gives the 0's in the $ab$, $ba$ and $bb$ cells will also give a 0 for the $+-$ entry in the $aa$ cell (cf.\ the ticket in Fig.\ \ref{HU-state-raffle-ticket-design}). The $+-$ entry in the $aa$ cell of the correlation array for this Hardy-Unruh state becomes the $++$ entry in the $bb$ cell of the correlation array for the corresponding Hardy state. This is the entry that prevents us from simulating the correlation array in Fig.\ \ref{H-correlation-array} for this Hardy state. Any raffle that gives the 0's in the $aa$, $ab$ and $ba$ cells must give a 0 for the $++$ entry in the $bb$ cell (cf.\ the ticket in Fig.\ \ref{H-state-raffle-ticket-design}). 

\subsection{Hardy-Unruh states between maximally entangled and product states} \label{Hardy-Unruh states between maximally entangled and product states}

From the correlation array in Fig.\ \ref{HU-correlation-array} we can read off that 
\begin{equation}
\frac{\pr(+\!-\!|aa)}{\pr(+\!+\!|aa)} = \tan^2{\!\alpha}.
\end{equation}
As $\alpha$ approaches $\pi/2$, this ratio grows without bound and the clash with basic logic becomes particularly severe. The chain of conditionals $A_{a_+} \rightarrow B_{b_+} \rightarrow A_{b_+} \rightarrow B_{a_+}$ suggests that if Alice and Bob both peel $a$ and Alice finds +, Bob should find + as well. Yet, for that peeling combination and for $\alpha$ close to $\pi/2$, Bob will almost always find $-$ instead!

If $\alpha = \pi/2$, Eqs.\ (\ref{HU state bb})--(\ref{HU state aa}) for $\ket{\psi_{HU}(\alpha)}$ reduce to:
\begin{equation}
\ket{\psi_{HU}(\textstyle{\frac\pi2})} = -\ket{+-}_{bb} = \ket{++}_{ba} = - \ket{--}_{ab} = \ket{-+}_{aa},
\label{HU product state}
\end{equation}
which is a product state. So we have the paradoxical situation that the clash with ordinary logic gets \emph{worse} as $\alpha$ approaches $\pi/2$ but \emph{disappears} when $\alpha = \pi/2$!\footnote{As \citet[p.\ 4]{Unruh:2018} observes: ``the closer the state is to a product state, a completely un-entangled state, the lower is the probability that if $A$ then $D$'' (where Unruh's $A$ and $D$ are our $A_{a_+}$ and $B_{a_+}$).\label{Unruh quote}} On closer inspection, this discontinuity is only apparent. From the correlation array in Fig.\ \ref{HU-correlation-array} we read off that 
\begin{equation}
\pr(+\!-\!|aa) = \frac{\cos^4{\!\alpha}\sin^2{\!\alpha}}{1 + \cos^2{\!\alpha}}.
\label{troublesome probability}
\end{equation}
This probability steadily decreases as $\alpha$ approaches $\pi/2$ and vanishes for $\alpha=\pi/2$. In fact, as both the correlation array in Fig.\ \ref{HU-correlation-array} and Eq.\ (\ref{HU product state}) show, if $\alpha$ approaches $\pi/2$ and Alice and Bob both peel $a$, the outcome is almost certainly $-+$.

Inspection of Fig.\ \ref{Hardy-Unruh-bananas} tells us that, if $\alpha = \pi/2$, $| +\rangle_b = |-\rangle_a$ and $|- \rangle_b = - |+\rangle_a$ (this explains the minus signs in Eq.\ (\ref{HU product state})). In other words, the operators representing `taste when peeled $a$' and `taste when peeled $b$' have the same set of eigenvectors. These operators thus commute, in which case the correlations found in measurements on this state can easily be simulated classically (e.g., with one of our raffles). What this means physically becomes clear if we substitute spin-$\frac12$ particles for our bananas for a moment. If $\varphi_{ab} = 2 \alpha = \pi$,  the directions $a$ and $b$ are exactly opposite to one another. Spin up/down in the $a$ direction then becomes spin down/up in the $b$-direction. The operators representing those observables obviously commute. In fact, we get from one to the other simply by relabeling eigenvectors and eigenvalues.

Since $\ket{\psi_{HU}(\textstyle{\frac\pi2})} = \ket{-+}_{aa} = \ket{--}_{ab}$ (see Eq.\ (\ref{HU product state})), it is impossible for Alice to peel $a$ and find $+$ if $\alpha = \pi/2$. This can also be read off the correlation array in Fig.\ \ref{HU-correlation-array}: all entries in the first row vanish for $\alpha = \pi/2$, which means that $\pr(A_{a_+}) =0$.  Hence, for $\alpha = \pi/2$, there is (1) no broken arrow and (2) no problem designing a raffle to simulate the quantum correlations:
\begin{enumerate}
\item Since its antecedent is false, the conditional $A_{a_+} \rightarrow B_{a_+}$ is vacuously true and perfectly compatible with the chain of conditionals $A_{a_+} \rightarrow B_{b_+} \rightarrow A_{b_+} \rightarrow B_{a_+}$ in Eq.\ (\ref{chain of conditionals}).
\item Our raffle will have no tickets with $+$ for $a$ on Alice's side, so we avoid the problem with the design of tickets brought out in Fig.\ \ref{H-state-raffle-ticket-design}.   
\end{enumerate}

If $\alpha=0$, Eqs.\ (\ref{HU state bb}) and (\ref{HU state aa}) for the Hardy-Unruh state $\ket{\psi_{HU}(\alpha)}$ reduce to
\begin{equation}
\ket{\psi_{HU}(0)} =  \frac{1}{\sqrt{2}}  \Big( \ket{+-}_{bb} -  \ket{-+}_{bb} \Big) =  \frac{1}{\sqrt{2}}  \Big( \ket{+-}_{aa} -  \ket{-+}_{aa} \Big),
\label{HU state 0 = singlet state}
\end{equation}
which is just the maximally entangled singlet state in Eq.\ (\ref{singlet state}). Yet, there is no clash with basic logic and no problem simulating the experiment with one of our raffles. As in the case of the Hardy state $\ket{\psi_{H}(\alpha)}$, which becomes maximally entangled if $\alpha = \pi/2$ (see Eq.\ (\ref{Hardy pi/2 max entangled})), this is because the peeling directions $a$ and $b$ coincide if $\alpha = \varphi_{ab}/2 =0$.  And the tastes of pairs of bananas in the singlet state only exhibit correlations that we cannot simulate with any of our raffles if Alice and Bob get to choose between \emph{different} peeling directions.

\section{Geometrical representation of the correlations found with Hardy-Unruh states} \label{Geometrical representation of the correlations found with Hardy-Unruh states}

What can we say about the local polytope $\mathcal{L}$ and the quantum convex set $\mathcal{Q}$ for the Hardy-Unruh setup (cf.\ Fig.\ \ref{elliptope})? 

\begin{figure}[h]
\centering
    \includegraphics[width=3.8in]{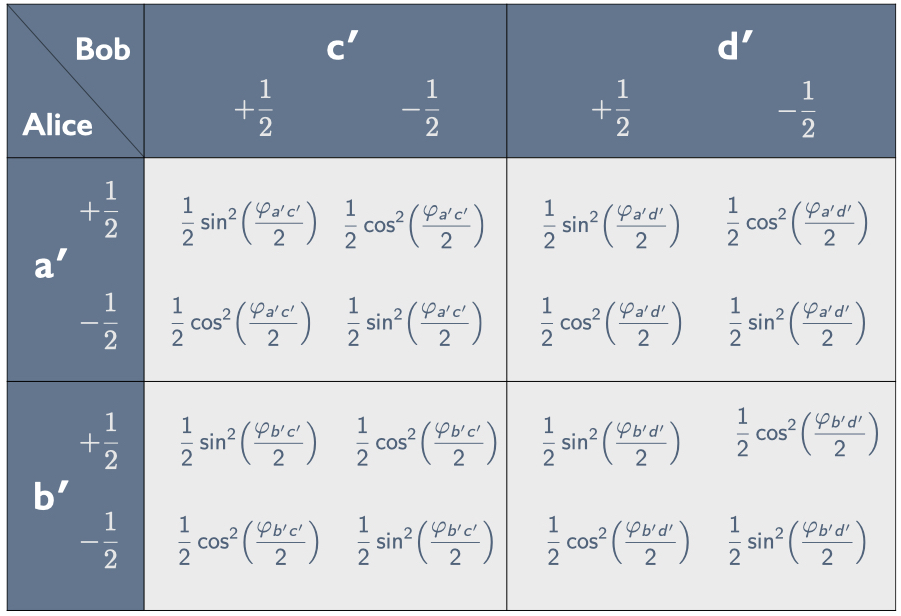}
 \caption{\small{Correlation array for the tastes $\pm\frac12$ of pairs of bananas in the singlet state (see Eq.\ (\ref{singlet state})), one of them peeled $a'$ or $b'$ by Alice, the other peeled $c'$ and $d'$ by Bob.}}
 \label{HU-correlation-array-symmetrized}  
\end{figure}

To answer this question, we start by comparing the correlation array in Fig.\ \ref{HU-correlation-array} for the tastes of pairs bananas, peeled $a$ or $b$, in the state 
$\ket{\psi_{HU}(\alpha)}$ in Eqs.\ (\ref{HU state bb})--(\ref{HU state aa}) (the Hardy-Unruh setup) to the correlation array in Fig.\ \ref{HU-correlation-array-symmetrized} for the tastes of pairs bananas, one peeled $a'$ or $b'$, the other peeled $c'$ or $d'$, in the state $\ket{\psi_{\mathrm{singlet}}}$ in Eq.\ (\ref{singlet state}) (the CHSH setup). 

The correlation array for the CHSH setup consists of four cells of the form shown in Fig.\ \ref{CA-singlet-cell} and can be fully characterized by four correlation coefficients (see Eq.\ (\ref{chi = - cos})):
\begin{equation}
\begin{array}{ccc}
\chi_{a'c'} = - \cos{\varphi_{a'c'}},  &\quad& \chi_{a'd'} = - \cos{\varphi_{a'd'}},     \\[.4cm]
\chi_{b'c'} = - \cos{\varphi_{b'c'}},  && \chi_{b'd'} = - \cos{\varphi_{b'd'}}.  
\end{array}
\label{chi's for a'b'c'd'}
\end{equation}

The local polytope for this setup is given by the CHSH inequality and three similar pairs of inequalities \citep[pp.\ 160--161, Eqs.\ (5.4)--(5.7)]{JanasCuffaroJanssen:2022}:
\begin{equation}
\begin{array}{c}
-2 \; \le \; \chi_{a^\prime c^\prime}  \, + \, \chi_{a^\prime d^\prime} \, + \, \chi_{b^\prime c^\prime} \, - \, \chi_{b^\prime d^\prime}  \;  \le 2, \\[.2cm]
-2 \; \le \; -\chi_{a^\prime c^\prime}  \, + \, \chi_{a^\prime d^\prime} \, - \, \chi_{b^\prime c^\prime} \, - \, \chi_{b^\prime d^\prime} \;  \le 2,\\[.2cm] 
 -2 \; \le \; \chi_{a^\prime c^\prime}  \, - \, \chi_{a^\prime d^\prime} \, - \, \chi_{b^\prime c^\prime} \, - \, \chi_{b^\prime d^\prime}  \; \le 2, \\[.2cm] 
 -2 \;  \le \; - \chi_{a^\prime c^\prime}  \, - \, \chi_{a^\prime d^\prime } \, + \, \chi_{b^\prime c^\prime } \, - \, \chi_{b^\prime d^\prime} \; \le 2.
 \label{CHSH ineqs} 
\end{array}
\end{equation}
These inequalities can be found in the same way as the pair in Eq.\ (\ref{Mermin inequality}) for the Mermin setup (ibid., pp.\ 157--159: Fig.\ 5.1 shows the raffle tickets for the CHSH setup, Table 5.1 lists the $\chi$ values for these tickets).

The quantum convex set for the CHSH setup is given by a non-linear inequality, first obtained by \citet{Landau:1988}, that follows from the straightforward generalization of the elliptope inequality in Eq.\ (\ref{elliptope inequality}) if Alice and Bob have four rather than three different peelings to choose from:
\begin{equation}
|\chi_{a'c'}\chi_{b'c'}-\chi_{a'd'}\chi_{b'd'}| \le \sqrt{1-\chi_{a'c'}^2}\sqrt{1-\chi_{b'c'}^2}+\sqrt{1-\chi_{a'd'}^2}\sqrt{1-\chi_{b'd'}^2}.
\label{3-elliptope}
\end{equation}
(ibid., p.\ 166, Eq.\ (5.30), with $a$, $b$, $a'$ and $b'$ relabeled $a'$, $b'$, $c'$ and $d'$).

To use these inequalities for the Hardy-Unruh setup we need to modify the setup somewhat. The problem is that Eqs.\ (\ref{CHSH ineqs}) and (\ref{3-elliptope}) are derived for \emph{balanced} variables, i.e., their two possible values are each other's opposite and equiprobable (see Section \ref{Preliminaries}). This guarantees that their expectation values vanish, which greatly simplifies the expressions for standard deviations and correlation coefficients (see Eqs.\ (\ref{def correlation coefficient}) and (\ref{variance})). While the variables measured by Alice and Bob in the Hardy-Unruh setup have opposite values, their expectation values do not vanish, as these two values are not equiprobable. 

We therefore introduce new variables that \emph{are} balanced but have the same covariances as the original ones. The correlations between these new balanced variables for a modified Hardy-Unruh setup can be simulated by a CHSH setup with appropriately chosen peeling directions.\footnote{Given a correlation array for  measurements on any two-particle state we can find a correlation array with the same correlation coefficients (though not the same expectation values) for measurements on the singlet state. This is a direct consequence of theorem 1 in \citet[pp.\ 93--94; for the proof, see Tsirelson, 1987]{Tsirelson:1980}. 
Our introduction of balanced variables for the Hardy-Unruh setup was inspired by the proof of part of Tsirelson's theorem in \citet[p.\ p.\ 7]{Avisetal:2009}.} Moreover, the modification preserves an important property of the correlation array for the Hardy-Unruh setup in Fig.\ \ref{HU-correlation-array}: the $ab$ and $ba$ cells are identical. Hence, we only need three $\chi$ parameters to characterize the correlation array for the CHSH setup with which we can simulate the correlations found in the modified Hardy-Unruh setup. This means that the local polytope and the quantum convex set for the modified Hardy-Unruh setup---like those for the Mermin setup (see Fig.\ \ref{elliptope})---can be pictured in three dimensions.

\begin{figure}[h]
\centering
    \includegraphics[width=5in]{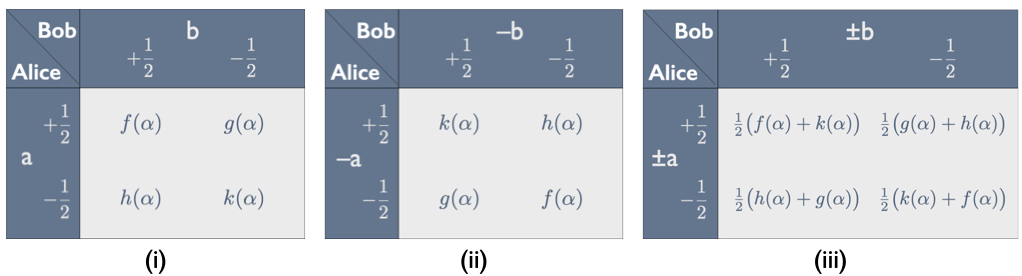}
 \caption{\small{Constructing balanced variables for the Hardy-Unruh setup. The figure shows a cell in the correlation array for Alice and Bob---peeling $a$ and $b$, respectively---recording (i) the tastes of their bananas, (ii) minus those tastes and (iii) the tastes in even and minus the tastes in odd runs. The functions $f(\alpha)$, $g(\alpha)$, $h(\alpha)$, $k(\alpha)$ can be read off the correlation array in Fig.\ \ref{HU-correlation-array}.}}
 \label{HU-setup-balanced}  
\end{figure}

We introduce the new balanced variables for the modified Hardy-Unruh setup in two steps. The three panels in Fig.\ \ref{HU-correlation-array} illustrate the process for the $ab$ cell. First, we imagine Alice and Bob, still choosing between peelings $a$ and $b$, recording the \emph{opposite} of the taste of their bananas. The correlation array for this experiment is obtained by switching the two entries on the diagonal and the two entries on the skew diagonal in each cell of the correlation array in Fig.\ \ref{HU-correlation-array} (see panel (ii) in Fig.\ \ref{HU-setup-balanced} for the $ab$ cell). This obviously flips the signs of the expectation values but does not affect the covariances. As we saw in Eq.\ (\ref{covariance formula}), in each cell, the covariance  is equal to $\frac14$ times the sum of the two entries on the diagonal minus $\frac14$ times the sum of the two entries on the skew diagonal. As these sums stay the same, so do the covariances. 

Next, we imagine Alice and Bob, still choosing between peelings $a$ and $b$, recording the taste of their bananas in even runs and the \emph{opposite} of the taste in odd runs. We obtain the correlation array for this experiment by taking, for all 16 entries, the straight average of the entries in the correlation arrays for the even and the odd runs (see panel (iii) in Fig.\ \ref{HU-setup-balanced} for the $ab$ cell). The four covariances are the same in all runs so the covariances for this combined correlation array will still be the same as for the original correlation array in Fig.\ \ref{HU-correlation-array}. But by having Alice and Bob alternate between recording the taste and recording minus the taste of their bananas, we ensure that the variables they measure are balanced. 

Panel (iii) in Fig.\ \ref{HU-setup-balanced} shows this for the $ab$ cell but it is true for all four cells of the combined correlation array. Both entries on the diagonal are the average of the two entries on the diagonal in the original correlation array and both entries on the skew diagonal are the average of the two entries on the skew diagonal in the original correlation array. Hence, in each cell, the sum of the two entries in each row and in each column gives $\frac12$ times the sum of all four probabilities in that cell. The entries in each row and in each column of each cell therefore sum to $\frac12$, which means that the variables measured by Alice and Bob when they alternate between recording the taste and minus the taste of their bananas are indeed balanced.

In each cell of the correlation array for the balanced Hardy-Unruh setup, as we will call it, the two entries on the diagonal and the two entries on the skew diagonal can be set equal to $\frac12$ times the square of, respectively, the sine and the cosine of some angle. Since two of its four cells are identical, the correlation array for  the balanced Hardy-Unruh setup can thus be fully characterized by three angles. Identifying these angles with half the angles $\varphi_{a'c'}$, $\varphi_{a'd'} = \varphi_{b'c'}$ and $\varphi_{b'd'}$ between the peeling directions $a'$, $b'$, $c'$ and $d'$, we can cast this correlation array in the form of the correlation array for the CHSH setup in Fig.\  \ref{HU-correlation-array-symmetrized}. The standard deviations for the variables in this correlation array are all $\frac14$, so the three correlation coefficients characterizing it are given by
\begin{eqnarray}
\begin{array}{l}
\chi_{a'c'}(\alpha) = 4 \braket{A_{a'} B_{c'}} = 4 \braket{A_a B_a}\!,   \\[.2cm]
\chi_{a'd'}(\alpha) = \chi_{b'c'}(\alpha) = 4 \braket{A_{a'} B_{d'}} = 4 \braket{A_{b'} B_{c'}} = 4 \braket{A_a B_b} = 4 \braket{A_b B_a}\!,  \\[.2cm]
\chi_{b'd'}(\alpha) = 4 \braket{A_{b'} B_{d'}} = 4 \braket{A_b B_b}\!.
\end{array}
\label{chis for CHSH proxy}
\end{eqnarray}
where we used that the covariances for this CHSH setup are the same as those for the original Hardy-Unruh setup. 

\begin{figure}[h]
\centering
    \includegraphics[width=4in]{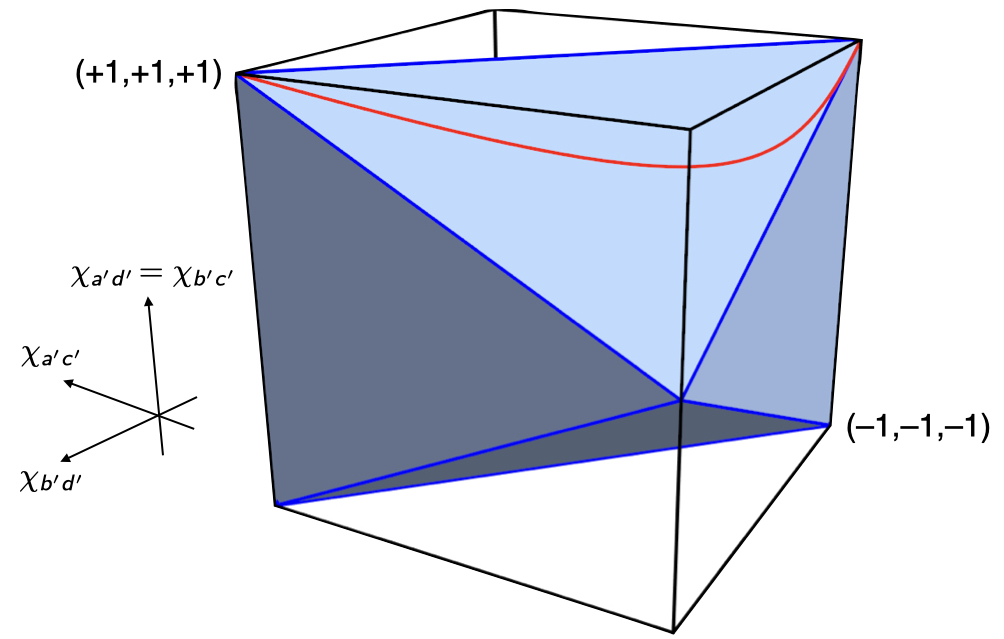}
 \caption{\small{Local polytope $\mathcal{L}$ for the CHSH setup with with two identical correlation coefficients. The red curve between two of the vertices of $\mathcal{L}$ represents the correlations found in the balanced Hardy-Unruh setup for states $\ket{\psi_{HU}(\alpha)}$ with $0 \le \alpha \le \frac\pi2$.}}
 \label{HU-L}  
\end{figure}

\begin{figure}[h]
\centering
    \includegraphics[width=4in]{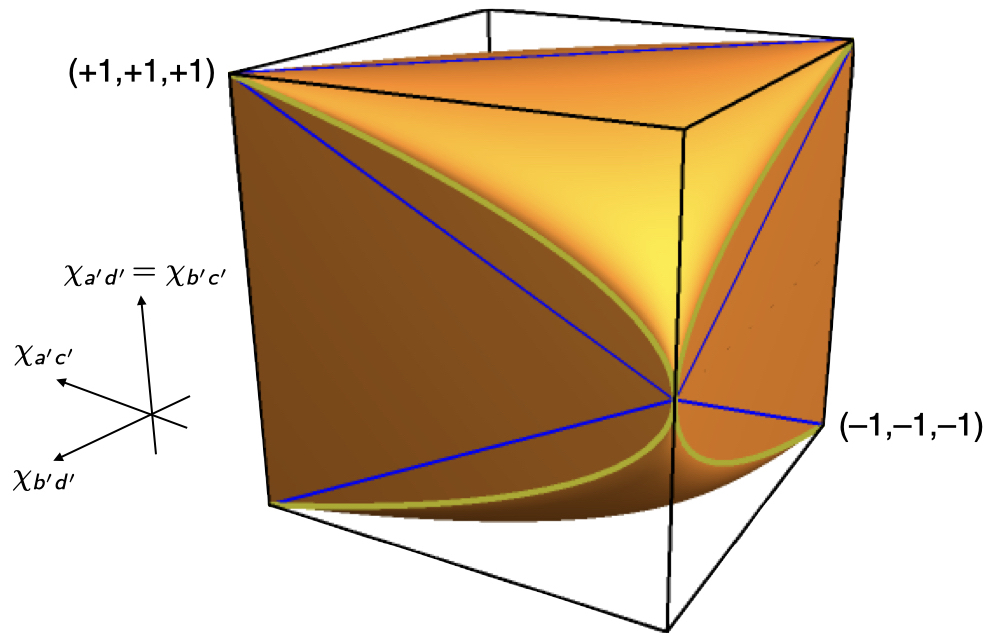}
 \caption{\small{Quantum convex set $\mathcal{Q}$ for the CHSH setup with two identical correlation coefficients.}}
 \label{HU-QandL}  
\end{figure}

Figs. \ref{HU-L} and \ref{HU-QandL} show the local polytope $\mathcal{L}$ and the quantum convex set $\mathcal{Q}$ for the subclass of correlations found in the CHSH setup if two of its four correlation coefficients are identical.\footnote{The depiction of the convex sets in Figs. \ref{HU-L} and \ref{HU-QandL} can also be found in \citet[p.\ 9, Fig.\ 2]{Leetal:2023}, who moreover extensively review the larger convex sets in the absence of the $\chi_{a'd'}=\chi_{b'c'}$ constraint.} We obtain the inequalities defining $\mathcal{L}$ and $\mathcal{Q}$ in this case by setting $\chi_{a'd'} = \chi_{b'c'}$ in Eqs.\ (\ref{CHSH ineqs}) and (\ref{3-elliptope}). We created Figs. \ref{HU-L} and \ref{HU-QandL} by feeding the resulting inequalities into Mathematica. Note the similarity of these figures to Fig.\ \ref{elliptope} for the Mermin setup. In both cases, $\mathcal{Q}$ looks like an inflated version of $\mathcal{L}$.\footnote{This `inflation' corresponds to the \emph{pushout} operation in \citet[pp.\ 10--11]{Leetal:2023} and was first found by \citet{Masanes:2003}.}
 
\begin{figure}[h]
\centering
    \includegraphics[width=5.5in]{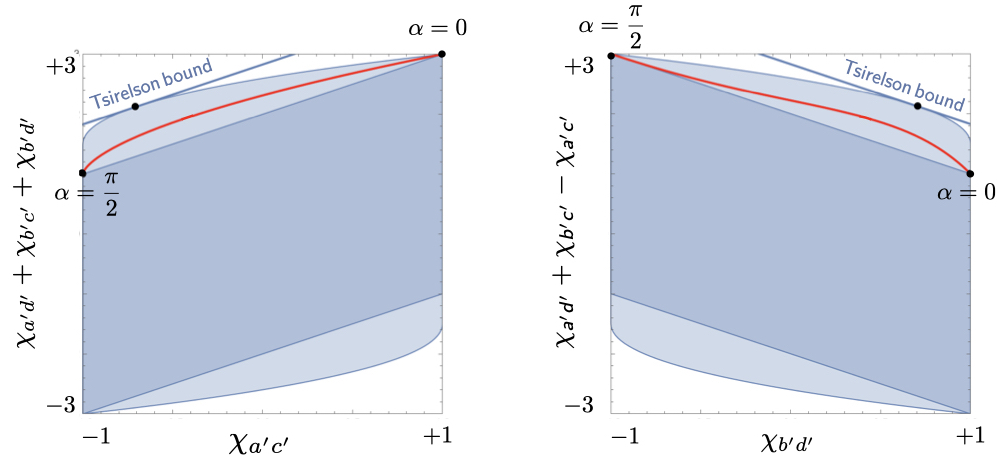}
 \caption{\small{Projections of the local polytope $\mathcal{L}$ (the dark blue parallelograms) and the quantum convex set $\mathcal{Q}$ ($\mathcal{L}$ plus the light blue extensions) for the CHSH setup with two identical correlation coefficients.  The red curve represents the projection onto the same plane of the curve representing the correlations found in the balanced Hardy-Unruh setup for the states $\ket{\psi_{HU}(\alpha)}$ in Eqs.\ (\ref{HU state bb})--(\ref{HU state aa}) with $0 \le \alpha \le \frac\pi2$.}}
 \label{HU-curve}  
\end{figure}

The values of the correlation coefficients in Eq.\ (\ref{chis for CHSH proxy}) parametrize the curve shown in Figs.\ \ref{HU-L} and \ref{HU-curve} representing the correlations found between the values of the balanced variables measured on the state $\ket{\psi_{HU}(\alpha)}$ for $0 \le \alpha \le \frac\pi2$ in our balanced Hardy-Unruh setup. We can compute the covariances on the right-hand side of Eq.\ (\ref{chis for CHSH proxy}) for these correlation coefficients with the help of the correlation array in Fig.\ \ref{HU-correlation-array} (cf.\ Eq,\ (\ref{covariance formula})):
\begin{equation}
\begin{array}{l}
\braket{A_a B_a} = {\displaystyle \frac14 \cdot \frac{2\cos^6{\!\alpha} - \sin^2{\!\alpha} \, (1 + \cos^2{\!\alpha})^2 - \cos^4{\!\alpha}\sin^2{\!\alpha}}{1 + \cos^2{\!\alpha}},}  
\\[.4cm]
\braket{A_a B_b} = \braket{A_b B_a} = {\displaystyle \frac14 \cdot \frac{\cos^4{\!\alpha} + 1- \cos^2{\!\alpha}\sin^2{\!\alpha}}{1 + \cos^2{\!\alpha}},}   
\\[.4cm]
\braket{A_b B_b} = {\displaystyle \frac14 \cdot \frac{2\cos^2{\!\alpha} - \sin^2{\!\alpha}}{1 + \cos^2{\!\alpha}}.}   
\end{array}
\label{chi(alpha)}
\end{equation}

Multiplying these expressions by 4 and feeding them into Mathematica, we found the curve in Figs.\ \ref{HU-L} and \ref{HU-curve}.\footnote{Both projection plots in Fig. \ref{HU-curve} are in the spirit of \citet{Gohetal:2018}. The convex sets depicted are equivalent (up to minor change of variables) to those in Fig.\ 5 on p.\ 8 of their paper. This is not entirely trivial: their Fig.\ 5 depicts the $\chi_{a'd'}=\chi_{b'c'}=\chi_{b'd'}$ \emph{section} of the relevant convex sets, whereas for instance our first plot amounts to a \emph{projection} onto this same hyperplane. But the non-trivial symmetries for these sets \citep[see][]{Leetal:2023}, in conjunction with their convexity, ensure that Goh et al.'s \emph{section} and our \emph{projection} are identical.} These figures clearly show that the correlations found with the state $\ket{\psi_{HU}(\alpha)}$ are outside the local polytope. As one readily verifies using Eq.\ (\ref{chi(alpha)}), they violate the third pair of CHSH-type inequalities in Eq.\ (\ref{CHSH ineqs}):  
\begin{equation}
 \chi_{a'c'} -  \chi_{a'd'}  -  \chi_{b'c'}  -  \chi_{b'd'}  = - \, 2 - \frac{4 \cos^4{\!\alpha} \sin^2{\!\alpha}}{1 + \cos^2{\!\alpha}}. 
 \label{CHSH violation and troublesome probability} 
\end{equation}
The second term on the right-hand side makes the left-hand side smaller than $-2$. Comparison with Eq.\ (\ref{troublesome probability}) shows that this term is equal to 4 times the probability $\pr(+\!-\!|aa)$ of the outcome responsible for the broken arrow found with the state $\ket{\psi_{HU}(\alpha)}$. As the following argument will show, this is no coincidence.

Let $A$ and $B$ represent the tastes found by Alice and Bob for some combination of peelings. Let  $\pr(\pm\pm)$ represent the probabilities of the four possible combinations of tastes. Solving four linear equations for these four probabilities, we can express them in terms of the expectation values and the covariance of $A$ and $B$.\footnote{These probabilities can also be expressed directly as expectation values of the corresponding operators in the quantum state under consideration. Take, for instance, the operators $\hat{A}_a$ and $\hat{B}_b$ representing the variables $A_a$, the taste of Alice's banana when peeled $a$, and $B_b$, the taste of Bob's banana when peeled $b$. We can write the projection operators onto the eigenvectors $\ket{\pm}_a$ and $\ket{\pm}_b$ of these operators as:
$$
\hat{P}_{a_\pm} =\frac{\hat{1} \pm 2\hat{A}_a}{2},\quad \hat{P}_{b_\pm} =\frac{\hat{1} \pm 2\hat{B}_b}{2}.
$$ 
For any ensemble of pairs of bananas (characterized by some density operator $\hat{\rho}$), the probability $\pr(+\!-\!|ab)$  is given by the expectation value of the tensor product of these projection operators: 
$$\pr(+\!-\!|ab) = \langle\hat{P}_{1_{a_+}} \otimes \hat{P}_{2_{b_-}} \rangle = {\textstyle \frac14} \, \langle  (\hat{1}_1 + 2\hat{A}_a) \otimes (\hat{1}_2 - 2\hat{B}_b)\rangle= {\textstyle \frac14} +  {\textstyle \frac12} \braket{\hat{A}_a}  - {\textstyle \frac12} \braket{\hat{B}_b} - \braket{\hat{A}_a \hat{B}_b},
$$
which corresponds to the second line in Eq.\ (\ref{pr in terms of exp and cov}).} The normalization of these probabilities gives us the first of these four equations:
\begin{equation}
\pr(++)+\pr(+-)+\pr(-+)+\pr(--) = 1;
\label{lin eqs for probs 1}
\end{equation}
the expectation values of $A$ and $B$ give us the second and the third:
\begin{eqnarray}
\langle A\rangle &\!\!\!=\!\!\!& {\textstyle \frac12} \big(\pr(++)+ \pr(+-) - \pr(-+) - \pr(--)\big), \label{lin eqs for probs 2}  \\[.2cm]
\langle B\rangle &\!\!\!=\!\!\!& {\textstyle \frac12} \big(\pr(++) - \pr(+-) + \pr(-+) - \pr(--) \big); \label{lin eqs for probs 3}
\end{eqnarray}
and the covariance of $A$ and $B$ gives us the fourth:
\begin{equation}
\langle AB\rangle = {\textstyle \frac14} \big( \pr(++) - \pr(+-) - \pr(-+) +\pr(--) \big).
\label{lin eqs for probs 4}
\end{equation}
Multiplying Eqs.\ (\ref{lin eqs for probs 2})--(\ref{lin eqs for probs 3}) by 2 and Eq.\ (\ref{lin eqs for probs 4}) by 4 and solving the resulting equations for the four probabilities, we find:
\begin{equation}
\begin{array}{ccc}
\pr(++)  &\!\!\!=\!\!\!&  \frac14 +  \frac12\! \braket{A} +  \frac12 \! \braket{B} + \braket{AB},  \\[.2cm]
\pr(+-)  &\!\!\!=\!\!\!&  \frac14 +  \frac12\! \braket{A} -  \frac12\! \braket{B} - \braket{AB},   \\[.2cm]
\pr(-+)  &\!\!\!=\!\!\!& \frac14 -  \frac12\! \braket{A} +  \frac12\! \braket{B} - \braket{AB},   \\[.2cm]
\pr(--)  &\!\!\!=\!\!\!&   \frac14 - \frac12\!  \braket{A} - \frac12\! \braket{B} + \braket{AB}.
\end{array}
\label{pr in terms of exp and cov}
\end{equation}

Now consider the probabilities that are $0$ in the $ab$, $ba$ and $bb$ cells of the correlation array in Fig.\ \ref{HU-correlation-array} and the non-vanishing probability in the $aa$ cell that is responsible for the broken arrow in the Hardy-Unruh chain. This gives us the following four equations:
\begin{equation}
\begin{array}{r}
\pr(+\!-\!|aa)  =  \frac14 +  \frac12\! \braket{A_a} -  \frac12\! \braket{B_a}- \braket{A_a B_a},  \\[.2cm]
0 =\pr(+\!-\!|ab)  =  \frac14 + \frac12\! \braket{A_a} - \frac12\! \braket{B_b} -\braket{A_a B_b}, \\[.2cm]
0 = \pr(+\!-\!|ba)  = \frac14 + \frac12\! \braket{A_b}- \frac12\! \braket{B_a} -\braket{A_b B_a}, \\[.2cm]
0=  \pr(-\!+\!|bb)  =  \frac14 - \frac12\! \braket{A_b} + \frac12\! \braket{B_b} -\braket{A_b B_b}. 
\end{array}
\label{special probs in HU corr array}
\end{equation}
If the last three are subtracted from the first, the expectation values all cancel and we are left with:
\begin{equation}
 \pr(+\!-\!|aa)  =  - {\textstyle \frac12} - \braket{A_a B_a} + \braket{A_a B_b} + \braket{A_b B_a} + \braket{A_a B_b}.
\end{equation}
Multiplying both sides by 4 and regrouping terms, we can rewrite this as:
\begin{equation}
 4\braket{A_a B_a} - 4\braket{A_a B_b} - 4\braket{A_b B_a}- 4\braket{A_a B_b} = - 2 - 4\pr(+\!-\!|aa) 
 \label{Bell inequality violation = broken arrow}
\end{equation}
Using Eq.\ (\ref{chis for CHSH proxy}) to replace 4 times the covariances by the corresponding correlation coefficients and using Eq.\ (\ref{troublesome probability}) for $\pr(+\!-\!|aa)$, we recover Eq.\ (\ref{CHSH violation and troublesome probability}). This shows, to reiterate, that the violation of the corresponding CHSH-type inequality is given by the probability of the outcome responsible for the broken arrow in the Hardy-Unruh chain. The maximum value of this probability is the same as the maximum value of the probability $\pr(+\!+\!|bb)$ of the outcome responsible for the broken arrow in the Hardy chain (see Eq.\ (\ref{maximum violation})).

\section{Conclusion} \label{Conclusion}

Our discussion of Hardy-Unruh chains has left us with a trifecta of deflating insights. The first of the three is that we cannot claim great originality for the other two. However, even those for whom the remaining two are hardly new will agree, we hope, that our use of the framework of \citet{JanasCuffaroJanssen:2022} has helped to put them in sharper relief. In this short concluding section, we summarize how our analysis in terms of raffle tickets, correlation arrays and their geometrical representation has led us to these insights.

The first insight is that the non-maximally entangled states giving rise to the broken arrow in Hardy's chain of conditionals in Eqs.\ (\ref{Hardy chain 12}) and (\ref{combined conditional}) are no different from those giving rise to the broken arrow in Unruh's chain of conditionals in Eqs.\ (\ref{Unruh chain 12}) and (\ref{chain of conditionals}). All these states are part of one large family (how large can be gleaned from our construction of a generic member in Eqs.\ (\ref{HU state bb'})--(\ref{HU state ab' basis partial})). We exhibited these family ties by constructing correlation arrays for correlations leading to both kinds of broken arrows, the one in Fig.\ \ref{H-correlation-array} for the Hardy states $\ket{\psi_{H}(\alpha)}$ in Eqs.\ (\ref{H state aa basis}) and (\ref{H state bb basis})--(\ref{H state ab basis}), the one in Fig.\ \ref{HU-correlation-array} for the Hardy-Unruh states  $\ket{\psi_{HU}(\alpha)}$ in Eqs.\ (\ref{HU state bb})--(\ref{HU state aa}). We showed how the defining properties of Hardy and Hardy-Unruh chains of conditionals can be read off these correlation arrays. We then showed that these two correlation arrays differ only in how they are labeled (the peelings $a$ and $b$, the tastes $+$ and $-$, and the angle $\alpha$ parametrizing the states). Although we only did this for members of a specific branch of the Hardy-Unruh family, it is clear that the same could be done for any family member.

The second insight is that a broken arrow in a Hardy-Unruh chain is equivalent to the violation of some Bell inequality. We showed this (again: for a special branch of the Hardy-Unruh family) by constructing a geometrical representation of the correlation array for the Hardy-Unruh setup (see Figs.\ \ref{HU-L}, \ref{HU-QandL} and \ref{HU-curve}). What complicated this task is that the two possible values of the variables measured in the Hardy-Unruh setup are not equiprobable. We took care of this problem by slightly modifying the Hardy-Unruh setup. We could then use a special case of the CHSH inequality (and similar inequalities associated with other facets of the local polytope) to characterize the class of correlations in this modified Hardy-Unruh setup allowed by a local hidden-variable theory (i.e., the class of correlations in this setup that can be simulated with one of our raffles). We showed (see Eq.\ (\ref{Bell inequality violation = broken arrow})) that the violation of one of these CHSH-type inequalities is given by the probability of the very outcome that is responsible for the broken arrow in the corresponding Hardy-Unruh chain. 

We agree with \citet[pp.\ 883--884]{Mermin:1994} that one should not exaggerate the difference between using one single outcome or the statistics of many outcomes as evidence that a correlation is not to be had in a local hidden-variable theory. If, for instance, we want to simulate the correlation array for a Hardy-Unruh setup in Fig.\ \ref{H-correlation-array} or \ref{HU-correlation-array} with one of our raffles, the problem is \emph{not} to get a non-zero probability for one particular outcome, but to get it \emph{while at the same time getting zero probabilities for several other outcomes}. In other words, rather than focusing on individual entries, we need to consider a correlation array as a whole.\footnote{For another (admittedly more convoluted) example involving a pair of spin-1 particles in the singlet state, see \citet[pp.\ 136--137]{JanasCuffaroJanssen:2022}.}

Despite being taken down a notch, Hardy-Unruh chains remain valuable. Whereas we usually consider violations of Bell inequalities by correlations found in measurements on maximally entangled states, Hardy-Unruh chains forcefully demonstrate that the slightest amount of entanglement  (cf.\ note \ref{Unruh quote}) already makes it impossible to simultaneously assign definite values to variables represented by non-commuting operators.

\section*{Acknowledgments}

Thanks to Jeff Bub, Mike Cuffaro and, especially, Bill Unruh for helpful discussion.

\end{document}